\documentclass[11pt,a4paper]{article}
\usepackage[T1]{fontenc}
\usepackage[latin1,utf8]{inputenc}

\usepackage{cancel}

\usepackage{amsmath,amssymb,amsfonts,multicol}
\usepackage[normal,font=small,labelfont=bf,labelsep=period]{caption}
\usepackage[pdftex]{color,graphicx}
\usepackage[english]{babel}
\usepackage[compress]{cite}

\usepackage[dvipsnames]{xcolor}
\usepackage{slashed}

\usepackage{enumitem}

\usepackage[linktoc=all,colorlinks=true,linkcolor=blue,urlcolor=magenta,citecolor=blue]{hyperref}
\hypersetup{%
  colorlinks = true,
  linkcolor  = black
}
\usepackage{cite}

\newcommand{\be}{\begin{equation}}
\newcommand{\ee}{\end{equation}}
\newcommand{\bea}{\begin{eqnarray}}
\newcommand{\eea}{\end{eqnarray}}

\newcommand{\eq}[1]{(\ref{#1})}

\definecolor{darkblue}{rgb}{0,0.2,0.6}
\definecolor{darkgreen}{rgb}{0,0.65,0}

\definecolor{gbcolor}{rgb}{.8,.3,.1}

\definecolor{mtcolor}{rgb}{.3,.2,.9}

\definecolor{corr}{rgb}{1,0,.2}

\numberwithin{equation}{section}

\begin{document}
\begin{flushright}
\footnotesize
{IFT-UAM/CSIC-17-084}\\
\end{flushright}
\color{black}
\vspace{0.3cm}

\begin{center}
{\huge \bf Primordial black hole dark matter\\ from single field inflation\\[1mm] }

\medskip
\bigskip\color{black}\vspace{0.6cm}

{
{\large\bf Guillermo Ballesteros}\ 
$^{a}$ and
{\large\bf Marco Taoso}
$^{b}$
}
\bigskip

{\it  $^a$ Institut de Physique Th\'eorique, Universit\'e Paris Saclay, CEA, CNRS}\\[1mm]
{\it $^b$  Instituto de F\'isica Te\'orica UAM/CSIC,\\Calle Nicol\'as Cabrera 13-15, Cantoblanco E-28049 Madrid, Spain.}\\[3mm]
\end{center}

\bigskip
\bigskip

\centerline{\large\bf Abstract}
\vspace{0.2cm}

We propose a model of inflation capable of generating a population of light black holes (about $10^{-16}$ -- $10^{-14}$ solar masses) that might account for a significant fraction of the dark matter in the Universe. The effective potential of the model features an approximate inflection point arising from two-loop order logarithmic corrections in well-motivated and perturbative particle physics examples. This feature decelerates the inflaton before the end of inflation, enhancing the primordial spectrum of scalar fluctuations and triggering efficient black hole production with a peaked mass distribution. At larger field values, inflation occurs thanks to a generic small coupling between the inflaton and the curvature of spacetime. We compute accurately the peak mass and abundance of the primordial black holes using the Press-Schechter and Mukhanov-Sasaki formalisms, showing that the slow-roll approximation fails to reproduce the correct results by orders of magnitude. We study as well a qualitatively similar implementation of the idea, where the approximate inflection point is due to competing terms in a generic polynomial potential. In both models, requiring a significant part of the dark matter abundance to be in the form of black holes implies a small blue scalar tilt with a sizable negative running and a tensor spectrum that may be detected by the next-generation probes of the cosmic microwave background. We also comment on previous works on the topic.

\begin{center}

\vfill\flushleft
\noindent\rule{6cm}{0.4pt}\\
{\small  E-mail addresses: \tt guillermo.ballesteros@cea.fr, m.taoso@csic.es}

\end{center}
\bigskip

\newpage
\tableofcontents

\section{Introduction}
\label{intro}

Soon after the first detection of gravitational waves (GW) emitted by a binary black hole (BH) merger \cite{Abbott:2016blz}, the possibility that BHs could constitute a significant amount of the Universe's dark matter (DM) started to regain attention swiftly. The three detections of this kind of event that have been reported so far by the LIGO Scientific Collaboration \cite{Abbott:2016blz,Abbott:2016nmj,Abbott:2017vtc} already suggest the existence of a large population of BHs with masses of the order of a few tens of $M_\odot$ ($1.99\times 10^{33}$ g). These values of mass\footnote{Large ($>30 M_\odot$) BH masses are often said to be somewhat higher than those that were expected to be detected first by LIGO according standard formation scenarios \cite{TheLIGOScientific:2016htt}. See however \cite{Mandel:2015qlu,Belczynski:2016obo,Rodriguez:2016kxx,Marchant:2016wow,Elbert:2017sbr}.} have fuelled the (old) idea \cite{1966AZh43758Z, Hawking:1971ei, Carr:1974nx} that long-lived BHs could have been produced during the very early stages of the life of the Universe, see \cite{Bird:2016dcv, Clesse:2016vqa, Sasaki:2016jop, Carr:2016drx}. In particular, these (primordial) black holes (PBHs) might have originated from large cosmological fluctuations produced during inflation. As the characteristic comoving wavelength, $k$, of such large primordial fluctuations becomes comparable to the Hubble scale (after inflation ends) they would collapse to produce BHs. These would then behave as non-baryonic cold DM throughout the subsequent evolution of the Universe.

\begin{figure}[t!]
\begin{center}
\includegraphics[width= 0.85 \textwidth]{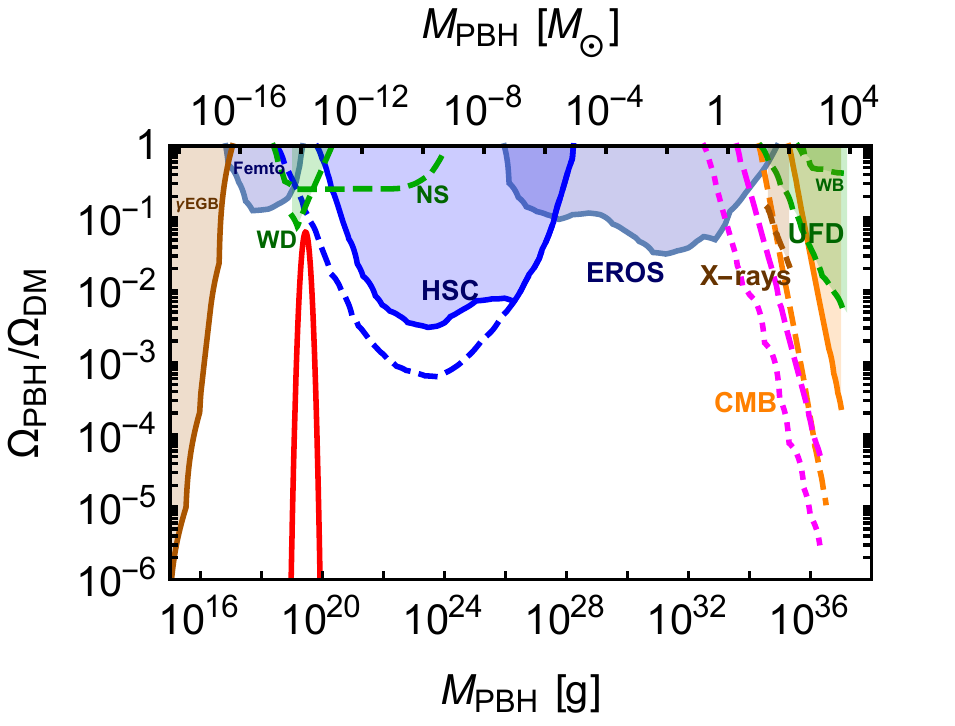} 
\caption{\em \small \label{fig:omega} {\bfseries } Fractional abundance of PBHs for the first example in table~\ref{tab:dataparameters} (red curve) and observational bounds (for a monochromatic mass spectrum).
The constraints are from measurements of the extragalactic gamma-ray background~\cite{Carr:2009jm}, femtolensing of gamma-ray bursts (Femto)~\cite{Barnacka:2012bm}, white dwarfs explosion (WD)~\cite{Graham:2015apa}, neutron star capture (NS)~\cite{Capela:2013yf}, microlensing  from Subaru (HSC)~\cite{Niikura:2017zjd} and
EROS/MACHO~\cite{Tisserand:2006zx},  wide binaries observations (WB)~\cite{Monroy-Rodriguez:2014ula}, dynamical heating of ultra-faint dwarf galaxies (UFD)~\cite{Brandt:2016aco,Koushiappas:2017chw} (we have taken the solid black line in figure 4 of~\cite{Brandt:2016aco}),  CMB measurements~\cite{Ali-Haimoud:2016mbv,Poulin:2017bwe} and radio and X-rays observations~\cite{Gaggero:2016dpq}.
{The solid (dashed) line shows the constraints from HSC taking into account (neglecting) the effects of finite source size on the event rate of microlensing (see figure 25 in~\cite{Niikura:2017zjd})}. For the (orange) constraints from CMB anisotropies we also show conservative (solid) and stronger (dashed) bounds (see figure 14 from~\cite{Ali-Haimoud:2016mbv}). The magenta lines refer to the CMB constraints derived in \cite{Poulin:2017bwe} assuming that PBHs form with an accretion disk (dashed and dotted lines refer respectively to the blue and red areas in their figure 4).
}
\end{center}
\end{figure}

The mass of each of the PBHs that are born in this way is inversely proportional to $k^2$, 
\begin{align} \label{massintro}
M(k)\simeq 5\,\gamma\times 10^{18} \,\left(\frac{k}{7\times 10^{13}\, {\rm Mpc}^{-1}}\right)^{-2}\,{\rm g}\,,
\end{align}
where $\gamma$ is a constant that models the efficiency of the process and can be analytically estimated to be around 0.2 \cite{Carr:1975qj}. Given that inflation scans through all the range of scales that are observable today, an interval of masses which spans many orders of magnitude is a priori possible for PBHs contributing to DM. {However, multiple bounds exist on this large range. The current situation is shown in figure \ref{fig:omega}, assuming a monochromatic mass spectrum, appropriate for the very narrow mass distributions that we will consider later.\footnote{For extended mass functions the allowed parameter space for PBHs appears to shrink further~\cite{Carr:2017jsz}, although studies tailored  to  the various constraints separately are still needed.}  We emphasize that all the bounds come with {various degrees} of uncertainty, related {in some cases} to assumptions about the astrophysical parameters involved in deriving each of them. The bounds that are particularly sensitive to astrophysical uncertainties have been indicated in the plot with dashed/dotted lines. 

The vast range of possible PBH masses} is limited from below due to Hawking radiation, since very light PBHs ($\lesssim 10^{-17} M_\odot$) would have entirely evaporated by now \cite{Hawking:1974rv}. At the large-mass end, several upper bounds on PBHs as DM exist. 
Specifically, the dynamics of a star cluster in Eridanus II \cite{Brandt:2016aco} and stars in other dwarf galaxies  \cite{Koushiappas:2017chw} disfavour the possibility that the DM of the Universe could be in the form of BHs of a few tens of $M_\odot$.
The same mass range is also constrained by radio and X-ray {obsrvations, since emissions} in these frequencies would be produced by the accretion of interstellar gas onto the PBHs~\cite{Gaggero:2016dpq}.
A severe upper bound comes from the non-observation of BH accretion effects on the Cosmic Microwave Background (CMB); see \cite{Ricotti:2007au} for an early analysis. This was used in late 2016  to exclude PBHs of masses $\gtrsim 10^2 M_\odot$ as the main component of the DM \cite{Ali-Haimoud:2016mbv}. Shortly after, it was argued in \cite{Blum:2016cjs} (using the same radial accretion modelling as in \cite{Ricotti:2007au}) that PBH DM with masses above $5 M_\odot$ is disfavoured.  Moreover, it has been pointed out very recently that if the gravitational collapse leading to PBHs occurs via an accretion disk (instead of respecting spherical symmetry), the CMB implies that PBHs more massive than 2$M_\odot$ cannot be the dominant component of the DM \cite{Poulin:2017bwe}. 
{Finally, the authors of \cite{Mediavilla:2017bok} have performed a study of gravitational microlensing using lensed images of 24 quasars.
The quasar microlensing events that they have identified are shown to be well fitted by a (monochromatic) distribution of compact objects in the lens galaxies with masses (in the approximate range of $0.05 M_\odot$--0.45$M_\odot$) and mass fraction in agreement with the expected values for the stellar component}. This has lead them to conclude that the the possibility of the DM being constituted by PBH with masses higher than $10 M_\odot$ is highly unlikely. Clearly, a conservative interpretation of the cosmological and astrophysical observations is required in order to make the case that a sizable fraction of the DM abundance could be explained by PBH of large mass. In this respect, it would be very important to {complement and extend with new data} the constraints from quasar microlensing \cite{Mediavilla:2017bok}, quantifying the amount of PBH DM that they allow in that range. Further studies on the constraining power of the CMB {would also be helpful.}

{There is however a different mass window (around $\sim 10^{-16.5}$--$10^{-13}\, M_\odot$)  for which a significant fraction of the DM {(perhaps most or even all of it in the range $\sim 10^{-14}-10^{-13}\, M_{\odot}$)} might be due to PBHs. The limits that are currently available in the literature for these PBHs come from neutron star capture in globular clusters (NS)~\cite{Capela:2013yf}, white dwarfs explosions (WD)~\cite{Graham:2015apa}, femtolensing of gamma-ray bursts (Femto)~\cite{Barnacka:2012bm} and microlensing  from Subaru (HSC)~\cite{Niikura:2017zjd}. {Taken at face value, these  bounds imply that monochromatic PBHs in this range can account at most for $\mathcal{O}(10)\%$ of the DM content. However, some of the astrophysical bounds constraining this region are arguably more uncertain than those cutting the large-mass end, such as the CMB ones. In particular, concerning the HSC constraints, since the Schwarzschild radii for light ($\lesssim10^{-10}$ $M_{\odot}$) PBHs becomes comparable 
or smaller than the wavelength of the HSC r-band filter, the corresponding wave-effect is expected to weaken the constraints for those masses (once it is computed, see the discussion in~\cite{Niikura:2017zjd}). If the NS and WD bounds could also be relaxed (which in principle is possible given the astrophysical uncertainties), PBHs of  $\sim 10^{-14}-10^{-13} M_\odot$ might be able to explain the totality of the DM}.} 
Although this low-mass region is not {directly relevant for} the recent LIGO detections of binary BH mergers, it is nonetheless important. {This is not only due to the uncertainties in the bounds, but also} because no known astrophysical mechanism can produce BHs so small.\footnote{It is worth putting in perspective the size of BHs of such small masses. The Schwarzschild radius of a BH is proportional to its mass. Knowing that it is about 3 km for a $\sim M_\odot$ BH, the typical scale of such a PBH candidate for DM is roughly between $0.1$ \AA\, and $1$ \AA, which is comparable to the size of the lightest atoms such as Hydrogen and Oxygen.} Therefore, postulating an alternative origin (such as primordial inflationary fluctuations) is  necessary if such objects are to be considered a possible, even subdominant, DM candidate. 

In this paper we investigate the extent to which single field models of inflation can produce a sizable amount of PBHs {that is interesting as a possible candidate for DM.} 
As mentioned above, the mass of the individual PBHs that are produced after inflation depends on the time at which the large fluctuations that seed their formation cross the Hubble scale. The times of Hubble crossing for a comoving mode of wavenumber $k$ are determined by the equation $k = \mathcal{H}\equiv {\dot{a}}/{a}$,
where $a$ is the scale factor of the Friedmann-Lema\^itre-Robertson-Walker (FLRW) metric of the Universe and $\mathcal{H}$ is the Hubble function defined with respect to conformal time. In the standard framework of inflation this happens twice for each $k$-mode; first during inflation itself, when $H =\mathcal{H}/a$ is approximately constant; and then after the end of it, when $H^{-1}$ grows in the subsequent epochs. In the usual single field slow-roll framework, the cosmological perturbations of wavenumber $k$ generated during inflation remain essentially constant in between these two crossing times, with an amplitude, $A_s$, that is approximately given by\footnote{We will later show that this approximation cannot be safely used in the cases of interest, and we will indeed require a more accurate expression. However, it is sufficient to illustrate well the point we want to convey now.}
\begin{align} \label{amp}
A_s=\frac{1}{24\pi^2 M_P^2} \frac{V}{\epsilon_V}\,,\quad  {\rm where} \quad \epsilon_V=\frac{M_P^2}{2}\left(\frac{V'}{V}\right)^2\,,
\end{align}
$V$ and $V'$ are the inflationary potential and its first derivative and $M_P=1/\sqrt{8\pi\,G}=2.8435\times 10^{18}$~GeV is the reduced Planck mass. Therefore, the mass of the PBHs is determined by the dynamics of the Universe during inflation and can be linked to the number of e-folds of expansion elapsed since the largest observable distance today became equal to $\mathcal H^{-1}$ during inflation. 

The CMB constrains $A_s$ to be of the order of $10^{-9}$ at those scales, whereas the values required for creating PBHs are much larger, typically $A_s\sim 10^{-3}$--$10^{-2}$. If we assume that the potential $V$ is nearly constant during inflation (which is indeed the case in standard slow-roll, leading to a quasi-de Sitter universe), the expression \eq{amp} tells that the required enhancement of $A_s$ may be achieved by significantly reducing the value of the slow-roll parameter $\epsilon_V$. Since this parameter quantifies the flatness of the potential, PBHs are produced provided that the rolling field encounters a sufficiently flat region of the potential during the course of inflation, which generates a peak in the spectrum of primordial fluctuations.  To the best of our knowledge, this idea was first proposed in \cite{Ivanov:1994pa}, where it was pointed out that a single-field inflationary potential that produces a PBH population capable of accounting for the DM must feature a near-inflection point. 

\begin{figure}[t]
\begin{center}
\includegraphics[width= 0.7 \textwidth]{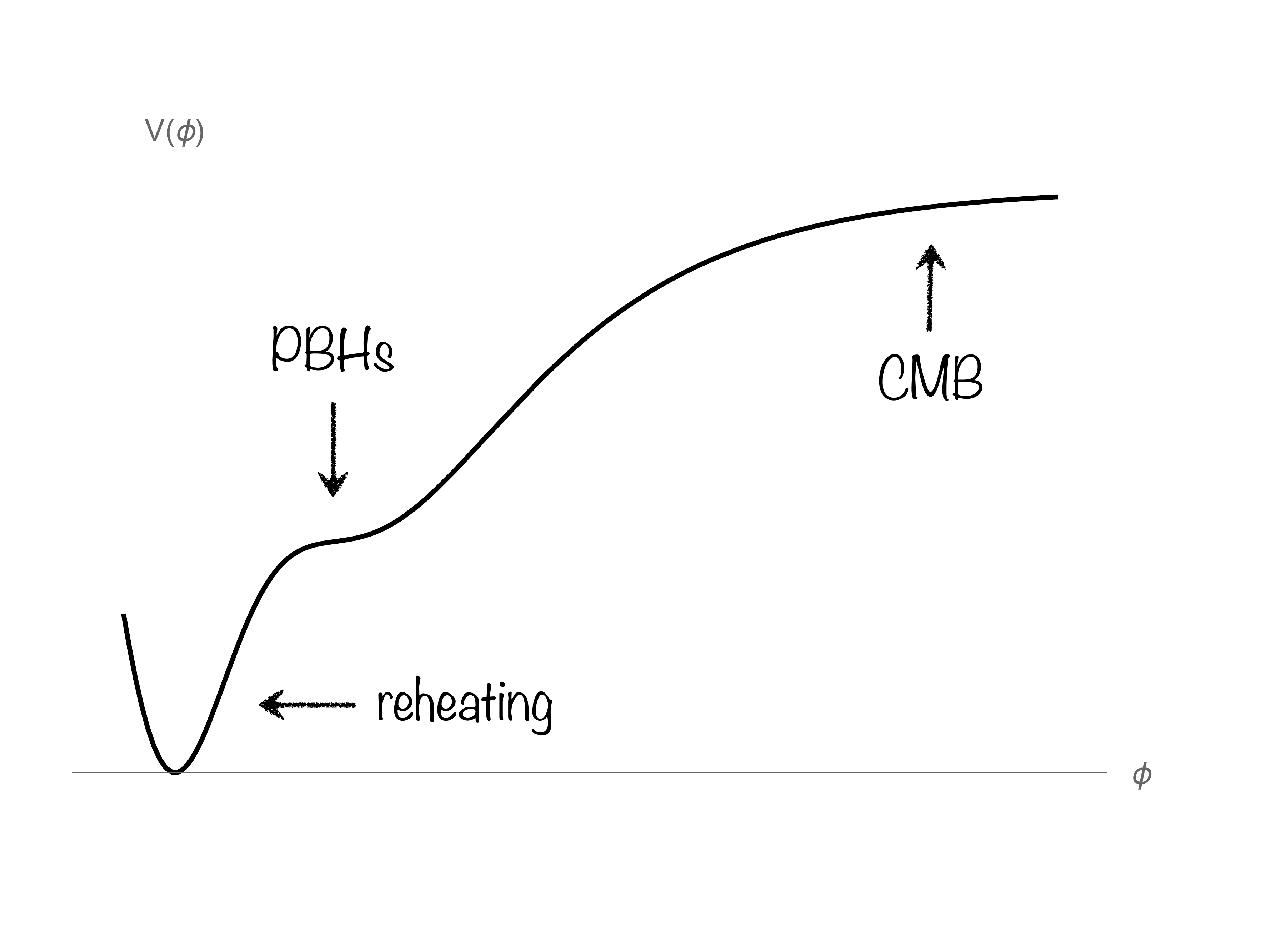} \vspace{-1cm}
\caption{\em \small \label{fig:scheme} {\bfseries } Schematic representation of the kind of inflationary potential required to fit the CMB data, produce PBHs and reheat the Universe after inflation.}
\end{center}
\end{figure}

A renormalizable potential that can have an inflection point is (see e.g.\ \cite{1990NuPhB.335..197H,Destri:2007pv}):
\begin{align} \label{V1}
V(\phi)=a_2\, \phi^2+a_3\, \phi^3+a	_4\,\phi^4\,,
\end{align}
where the $a_i$ can be considered constant. This potential vanishes at its absolute minimum in $\phi=0$, in agreement with the fact that the current cosmological constant is negligible in comparison with the energy density stored in the potential during inflation. 
A different possibility sharing these two features is a potential of the form \cite{Ballesteros:2015noa}:
\begin{align} \label{V2}
V(\phi)=\frac{\lambda_0}{4!}\left(1+q_1 \log\frac{\phi^2}{\phi_0^2}+q_2\left(\log\frac{\phi^2}{\phi_0^2}\right)^2+\cdots\right)\phi^4\,,
\end{align}
which describes radiative corrections to a renormalizable potential for large field values. In this case, $\lambda_0$, $\phi_0$ and the $q_i$ are assumed to be constants. Both types of potentials can be shown to be consistent with the current CMB constraints and to solve the horizon and flatness problems; see in particular \cite{Ballesteros:2015noa}. If their parameters are chosen adequately, these potentials can be sufficiently flat to reduce the amount of primordial gravitational waves with respect to the prediction of  $V(\phi)\propto\phi^4$, which is now ruled out by CMB data due to the strong Planck+BICEP2 constraint on the tensor-to-scalar ratio  \cite{Array:2015xqh}:
\begin{align} \label{rbound}
r< 0.07\quad {\rm at}\quad k=0.05\, {\rm Mpc}^{-1}\quad {\rm and}\quad  95\%\, {\rm C.L.}
\end{align}
The plateaus that these potentials can feature may instead be suitably engineered to generate PBHs, but in that case they do not fit the CMB data. This is because at large field values, away from the plateau involved in the PBH formation mechanism, the potentials \eq{V1} and \eq{V2} grow as $\phi^{4}$ and $(\log\phi)^2\,\phi^4$, respectively, causing an overproduction of primordial GW, which violates the upper bound \eq{rbound}. 

One could conjecture that the difficulty to obtain both PBHs and successful inflation could be surmounted by introducing a direct coupling between the scalar field $\phi$ and the curvature scalar $R$. For example, it is well-known that a $\phi^4$ potential leads to an asymptotic plateau at large field values --after redefining adequately the spacetime metric and the inflaton--  if the field $\phi$ is coupled to $R$ through a term of the form $\phi^2\, R$ \cite{Spokoiny:1984bd}. It might then be possible to generate PBHs {\it and} satisfy all the CMB constraints {\it and} obtain a sufficient amount of inflation from a potential such as \eq{V1} or \eq{V2} by including a non-negligible non-minimal coupling between $\phi$ and $R$.  After the appropriate field redefinitions, a potential with the required characteristics should schematically look like the one that is shown in figure \ref{fig:scheme}. The potential must be sufficiently flat for large values of the field (where the primordial spectrum observed with the CMB is generated) to satisfy the constraint \eq{rbound}. It must also have an approximate plateau at smaller field values (corresponding to larger values of $k$) to produce PBHs and, also, a minimum with $V=0$ to reheat the Universe after inflation ends. In this paper we explore if the PBHs produced by these models can be in the adequate mass range and have a sufficient abundance to constitute the DM of the Universe. {We find that imposing the Planck CMB constraints and requiring a reasonable number of e-folds of inflation, these models can generate a population of PBHs that falls in the low-mass window ($\sim 10^{-16.5}$--$10^{-13}\, M_\odot$) that is potentially interesting for DM.}

Before concluding this section let us mention that a toy potential with the qualitative features of figure \ref{fig:scheme} was recently put forward in \cite{Garcia-Bellido:2017mdw}. That potential is a ratio of polynomials which appears difficult to justify in an effective quantum field theory. We focus instead on standard renormalizable potentials (that grow as $\phi^4$ for large field values) and we emphasise the role of the non-minimal coupling to gravity to generate the inflationary plateau. We will comment further on that model in Section \ref{pbhPOL}. Besides, in reference \cite{Ezquiaga:2017fvi} a particular case of the potential \eq{V2} (corresponding to $q_1=0$) with non-minimal coupling to $R$ has been explored recently. We will comment on it at the end of Section \ref{pbhRP}. {Aside from the differences in the models themselves, our work goes beyond those analysis in the way we compute the primordial spectrum, for which we use the Mukhanov-Sasaki formalism \cite{Sasaki:1986hm,Mukhanov:1988jd}.}

In the next section we review the relevant formulae for the calculation of the mass and abundance of PBHs from the primordial spectrum of perturbations generated by inflation. In Section \ref{setup} we describe the inflationary set-up that we explore and provide the details of the method  that we use to compute the spectrum of primordial {perturbations.} In Section \ref{Numericalanalysis} we list the requirements that a model must satisfy for successful inflation and generating PBHs that may account for the DM of the Universe. In that section we also describe the strategy we follow to look for such models.  Then, in Sections \ref{pbhRP} and \ref{pbhPOL} we discuss our results for the potentials \eq{V2} and \eq{V1}, respectively. We present our conclusions in Section \ref{sec:conclusions}.

\section{Primordial black hole production}
\label{sec:abundance}

PBHs are formed when $\mathcal{H}$ becomes comparable to the wavelength of a sufficiently large primordial density fluctuation,  after inflation.  Their mass ($M$) is assumed to be directly proportional to the mass inside one Hubble volume at that time:
\begin{equation} \label{rel}
M=\gamma M_H =\gamma \frac{4}{3} \pi \rho H^{-3}\,,
\end{equation}
where the factor $\gamma$ depends on the details of the gravitational collapse.  The precise relation between $M$ and $M_H$ is uncertain.
Here we take $\gamma=0.2,$ as suggested by the analytical model described in \cite{Carr:1975qj} for PBHs formed during the radiation era, which is the situation we assume in what follows.
The relation between the comoving wavenumber, $k$, and the mass of the corresponding PBHs can be obtained using the conservation of entropy, $d(g_s(T)\, T^3\, a^3)/dt=0$, and  the scaling of the energy density, $\rho\propto g(T)\,T^4$, with the temperature, $T$, during the radiation era: 
\begin{equation}
M=\gamma\,M_{H(eq)}\,\left( \frac{g(T_{f})}{g(T_{eq})} \right)^{1/2} \left( \frac{g_s(T_{f})}{g_s(T_{eq})} \right)^{-2/3} \left( \frac{k}{k_{eq}} \right)^{-2}\,,
\end{equation} 
where $g(T)$  and $g_s(T)$ are the effective number of degrees of freedom in the radiation and the entropy densities, respectively; and the subscripts $_{eq}$ and $_f$ refer to the times of matter-radiation equality and PBH formation. The quantity  $M_{H(eq)}=4\pi\rho_{eq} H_{eq}^{-3}/3$ is the horizon mass at equality. Assuming $g(T)=g_s(T)$, which for our purposes is a good approximation even beyond electron-positron annihilation, one gets:
\begin{equation} \label{Mk}
M=10^{18}\mbox{ g }  \left( \frac{\gamma}{0.2}\right) \left( \frac{g(T_f)}{106.75}\right)^{-1/6} \left(\frac{k}{7 \times 10^{13} \mbox{ Mpc}^{-1}}\right)^{-2}\,,
\end{equation}
where we have used that $g(T_{eq})=3.38$, $k_{eq}=0.07\,\Omega_m\, h^2$ Mpc$^{-1}$ and we have written the result in terms of the Standard Model (SM) number of relativistic degrees of freedom deep in the radiation era, $g(T)=106.75$. Assuming the particle content of the SM, this expression then reduces to the formula \eq{massintro} of the Introduction.

In the context of the Press-Schechter model of gravitational collapse \cite{Press:1973iz}, the mass fraction in PBHs of mass $M$, which we denote $\beta(M)$, is given by the probability that the overdensity $\delta$  is above a certain threshold for collapse $\delta_c.$ Assuming that $\delta$ is a random gaussian variable with mass- (i.e.\ scale-) dependent variance, we have:
\begin{equation}\label{eq:beta}
\beta(M)  =  \frac{1}{\sqrt{2 \pi \sigma^2(M)}}\int_{\delta_c}^{\infty} d\delta\, \exp\left(\frac{-\delta^2}{2\sigma^2(M)}\right)\,.
\end{equation}
The shape of the probability distribution of $\delta(k)$ and the value of $\beta(M)$ (for a given $\delta_c$) are uniquely determined by the variance $\sigma^2(M)$, which we assume to be the coarse-grained variance of the density contrast smoothed on a scale $R=1/k$. For radiation domination this is given by:
\begin{equation}\label{eq:beta1}
\sigma^2(M(k))   =\int \frac{dq}{q} \mathcal{P}_{\delta}(q) W(q R)^2 =\frac{16}{81} \int \frac{dq}{q} \left( q R \right)^4  \mathcal{P}_{\mathcal{R}}(q) W(q R)^2\,,
\end{equation}
where $\mathcal{P}_{\mathcal{R}}$ and $\mathcal{P}_{\delta}$ are the dimensionless power spectra of the primordial comoving curvature perturbations and density fluctuations, respectively; see e.g. \cite{Young:2014ana} and \cite{Liddle:2000cg}.
For the smoothing window function we take for convenience a Gaussian: $W(x)=\exp(-x^2/2)$, although other choices are possible and  $\beta(M)$ should only be weakly dependent on them.

Finally, the present abundance of PBHs is:\footnote{This ignores the effects of evaporation (which certainly can be neglected for the masses of interest), mass accretion and PBH merging.}
\begin{equation}\label{eq:beta2}
\Omega_{PBH}=\int \frac{dM}{M} \Omega_{PBH}(M)\,, 
\end{equation}
with \cite{Inomata:2017okj}
\begin{equation}\label{eq:beta3}
\frac{\Omega_{PBH}(M)}{\Omega_{DM}} =   \frac{\beta(M)}{8\times 10^{-16}} \left( \frac{\gamma}{0.2} \right)^{3/2} \left( \frac{g(T_f)}{106.75} \right)^{-1/4} \left( \frac{M}{10^{18}\mbox{ g}}\right)^{-1/2}
\end{equation}
and the total cold DM abundance is $\Omega_{DM}\simeq0.26$~\cite{Ade:2015xua}.

As it can be noticed from equation \eq{eq:beta}, the mass fraction of PBHs is exponentially sensitive to the critical density for collapse $\delta_c.$ In other words, small variations of $\delta_c$ or $\mathcal{P}_{\mathcal{R}}$ imply dramatic changes in the present abundance of PBHs. 
During radiation domination, the most recent analyses suggest $\delta_c \simeq 0.45$~\cite{Musco:2004ak,Musco:2008hv,Musco:2012au,Harada:2013epa}. For this value, PBHs can account for an $\mathcal{O}(1)$ fraction of the dark matter content of the Universe only if the primordial power spectrum of curvature fluctuations is enhanced at the scale of the PBHs to be around $\mathcal{P}_{\mathcal{R}}\simeq 10^{-2}$.\footnote{This value can be lower if the primordial spectrum is sufficiently non-Gaussian.} As we mentioned in the Introduction, this is about seven orders of magnitudes larger than that at the CMB scale $\mathcal{P}_{\mathcal{R}}\simeq 10^{-9}.$ In the coming sections we will discuss in quantitative detail how such a feature can arise in well-motivated models of single-field inflation. 

Before proceeding with that, it is useful to relate the PBH mass \eq{Mk} to the amount of expansion that the Universe undergoes during inflation in between the time that some fiducial scale $k_*$ exited the Hubble radius and the time of exit for the scale of PBH formation. In order to do so, we assume that the Hubble scale during inflation, $H_I$, is approximately constant. The amount of expansion we want to quantify is then 
\begin{align}
\Delta N_e^* =\log\frac{a_k}{a_*}=\log\frac{a_k\, H_I}{k_*}=\log\frac{a_f\, H_f}{k_*}\,,
\end{align}
where the subscript $_k$ indicates the time (during inflation) at which $k=a\,H=\mathcal{H}$. Assuming, as before, that the number of effective degrees of freedom in the entropy and energy densities are equal, we find that \cite{Motohashi:2017kbs}:
\begin{align}
\Delta N_e^*=-\frac{1}{2}\log \frac{M}{M_\odot}+\frac{1}{2}\log \gamma+\frac{1}{12}\log\frac{g(T)}{106.75}+\frac{1}{2}\log\frac{4.4\times 10^{24}\,\Omega_r H_0^2}{k_*^2}\,,
\end{align}
where the radiation density is  $\Omega_r\,h^2=4.18\times 10^{-5}$ and $H_0 \simeq 0.0007$~Mpc$^{-1}$. Choosing for concreteness $k_*=k_{0.05}=0.05$ Mpc$^{-1}$, which is the reference scale used by the Planck collaboration, we finally get:
\begin{align} \label{Nesimp}
\Delta N_e^*=18.37-\frac{1}{2}\log \frac{M}{M_\odot}\,.
\end{align}
For instance, PBHs with mass $M=10^{-14}\, M_{\odot}$ are associated to a comoving scale $k^{-1}$ which becomes larger than $\mathcal{H}^{-1}$ approximately 34.5 e-folds after $k_{0.05}$ does it.

\section{Inflation and the spectrum of primordial perturbations}
\label{setup}

We consider a real scalar field coupled to gravity in the so-called Jordan frame:
\begin{equation} \label{act}
\mathcal{S}   = \int d^4x\, \sqrt{-g}\, \left(  - \frac{1}{2}\left(M_P^2+\xi\,\phi^2\right) R + \frac{1}{2} g_{\mu\nu}\,\partial^{\mu}\phi\,\partial^{\nu}\phi -V(\phi)\right)\,.
\end{equation}
We will later choose $V$ to be equal to either \eq{V1} or \eq{V2}, but we keep it generic for the moment. It is worth recalling that the coupling $\xi$ is generated radiatively even if it is zero at some scale, and therefore it should be included unless it is assumed to be so small that its effect is negligible. Since we only consider potentials that vanish at $\phi=0$, the term $\xi\,\phi^2\, R$ does not contribute to the actual Planck mass squared today, which is just  $m_P^2=8\pi\,M_P^2$. This coupling between $\phi$ and $R$ can be recast into a non-canonical kinetic term for the scalar field by performing a Weyl transformation of the metric:
\begin{align}
\tilde{g}_{\mu\nu}(x)=\Omega^2[\phi(x)]\,g_{\mu\nu}(x)\quad {\rm with}\quad \Omega^2=1+\frac{\xi \phi^2}{M_P^2}\,.
\end{align}
This leads to
\begin{align}
\mathcal{S}   = \int d^4x \sqrt{-\tilde{g}} \left(  - \frac{M_P^2}{2} \tilde{R} + \frac{1}{2} K(\phi)\, \tilde{g}_{\mu\nu}\, \partial^{\mu}\phi\,\partial^{\nu}\phi -\tilde{V}(\phi) \right)\,,
\end{align}
where
\begin{equation}\label{potentialtilde}
K = \frac{1}{\Omega^2} +\frac{3 M_P^2}{2}\left(\frac{d \log \Omega^2}{d \phi}\right)^2\quad {\rm and} \quad \tilde{V} = \frac{V}{\Omega^4}\,.
\end{equation}
The kinetic term of the scalar field can be canonically normalized with the following field redefinition:
\begin{equation}\label{fieldredef}
\Omega^2 \frac{d\chi}{d\phi} = \sqrt{\Omega^2+\frac{3 M_P^2}{2}\left(\frac{d \Omega^2}{d \phi}\right)^2 }\,.
\end{equation}
In terms of the new field,  $\chi(\phi)$, and defining
\begin{align}
U(\chi)=\tilde{V}\left(\phi(\chi)\right)\,,
\end{align} 
the action reads
\begin{equation} \label{act2}
\mathcal{S} = \int d^4x \sqrt{-\tilde g} \left(  - \frac{M_P^2}{2}\, \tilde R+ \frac{1}{2} {\tilde g}_{\mu\nu}\, \partial^{\mu}\chi\,\partial^{\nu}\chi -U(\chi) \right)\,.
\end{equation}
This is the form of the action that we will use in later sections of the paper to compute the inflationary dynamics and the formation of PBHs from the potentials \eq{V1} and \eq{V2}. 

\subsection{The dynamics of the inflaton as a function of the number of e-folds}

The evolution of the field $\chi$ when $\tilde g_{\mu\nu}$ is chosen to be a flat FLRW metric can be conveniently described in terms of the number of e-folds elapsed from an initial (cosmic) time $t_i$, which is defined as the integral
\begin{align}
N_e(t)=\int_{t_i}^t H(t') dt'\,.
\end{align} 
Concretely, $\chi(N_e)$ obeys the following differential equation \cite{Ballesteros:2014yva},
\begin{equation}\label{evo}
\frac{d^2\chi}{dN_e^2} +3\frac{d\chi}{dN_e} -\frac{1}{2 M_P^2} \left( \frac{d\chi}{dN_e} \right)^3 + \left( 3 M_P^2 -\frac{1}{2} \left( \frac{d\chi}{dN_e} \right)^2 \right)\frac{d \log U}{d\chi}=0\,,
\end{equation}
where $\sqrt{2\epsilon_U}=M_P\, (d \log U/d\chi)$ encodes the effect of the shape of the potential on the trajectory of the field. We remark that the equation \eq{evo} is exact; being, in particular, free from the slow-roll approximation. This equation can be solved numerically for any smooth $\epsilon_U$ and it allows a precise determination of the number of e-folds  between some $t_i$ and the end of inflation; which happens when the condition $\epsilon_H=1$ is satisfied, where 
\begin{align} \label{eH}
\epsilon_H=\frac{1}{2M_P^2}\left(\frac{d\chi}{dN_e}\right)^2. 
\end{align}

\subsection{Slow-roll approximation}

We consider potentials that are flat enough to guarantee the application of the slow-roll approximation for the computation of the spectrum of primordial perturbations {\it at CMB scales}. In this limit, the scalar and tensor spectra at those scales can be accurately expressed as:
\begin{align} \label{custom}
\mathcal{P}_\mathcal{R}(k)=  A_s\left(\frac{k}{k_*}\right)^{n_s-1+\frac{\alpha}{2}\ln\frac{k}{k_*}+\cdots}\,,\quad 
\mathcal{P}_t(k) = A_t\left(\frac{k}{k_*}\right)^{n_t+\cdots}\,,
\end{align}
where the parameters $n_s$, $\alpha \equiv {d n_s}/{d \ln k}$, etc.\  are implicitly evaluated at a fiducial scale $k_*$, for instance $k_*=k_{0.05}\equiv 0.05$\,Mpc$^{-1}$. At leading  order in the slow-roll expansion, these parameters are simple functions of the inflationary potential $U$ and its derivatives with respect to $\chi$, which we denote with primes:
\begin{align} \label{index} 
n_s\simeq 1+ 2\,\eta_U-6\,\epsilon_U\,, \quad
\alpha\simeq -2\,\xi_U +16\,\epsilon_U\,\eta_U-24\,\epsilon_U^2\,,\quad
 A_s=\mathcal{P}_\mathcal{R}(k_*)\simeq \frac{U}{24\pi^2\epsilon_U\,M_P^4}
\end{align}
and
\begin{align} \label{tens}
 \quad A_t\simeq r A_s \simeq 16\epsilon_U\, A_s\,,\quad n_t\simeq-\frac{r}{8}\,,
\end{align}
where
\begin{align} \label{srU}
\epsilon_U =\frac{M_P^2}{2}\left(\frac{U'}{U}\right)^2\,,\quad \eta_U =M_P^2\frac{U''}{U}\,,\quad \xi_U = M_P^4\frac{U'U'''}{U^2}\,.
\end{align}

A more accurate and precise determination of the scalar power spectrum is obtained using the expression \cite{Stewart:1993bc}:
\begin{align} \label{PH}
\mathcal{P}_\mathcal{R}\simeq \frac{1}{8\pi^2 M_P^2} \frac{H^2}{\epsilon_H}\,,
\end{align}
where the Hubble function squared can be computed (via Friedmann's equation) as 
\begin{align} \label{fried}
H^2=\frac{U}{(3-\epsilon_H)M_P^2}\,.
\end{align}
In the limit in which $\epsilon_H\ll 3$, we have from the last expression that $U\simeq 3\,M_P^2\,H^2$. If, in addition, the condition $|\eta_H| \ll 3$ is also satisfied, where
\begin{align} \label{etaH}
\eta_H=\epsilon_H-\frac{1}{2}\frac{d\log \epsilon_H}{d N_e}\,,
\end{align}
then the slow-roll attractor
\begin{align} \label{att}
\frac{d\chi}{d N_e}\simeq -\sqrt\frac{2\epsilon_U\,U}{3H^2}\simeq-M_P\sqrt{2\epsilon_U}
\end{align} 
is a good approximation, which also implies that $\epsilon_H\simeq \epsilon_U$  and, therefore
\begin{align} \label{app}
\mathcal{P_R}\simeq \frac{V}{24\pi^2\epsilon_U\,M_P^4}\,,
\end{align}
in agreement with the expression for $A_s$ given in \eq{index}. The equation \eq{PH} expresses the power spectrum of primordial fluctuations as a function of the classical inflaton's trajectory, which is governed by the equation \eq{evo}. On the other hand, the approximation \eq{app} to the more accurate expression \eq{PH} is a function of the inflaton's value alone, and, as we have explained, it implicitly assumes that the trajectory satisfies the slow-roll approximation $\epsilon_H\ll3$ and $|\eta_H|\ll3$. Indeed, the slow-roll parameters can also be written as
\begin{align}
\epsilon_H=\frac{\dot\chi^2}{2M_P^2 H^2}\,,\quad \eta_H=-\frac{\ddot\chi}{H\dot\chi}\,,
\end{align}
where the dots indicate derivatives with respect to cosmic time. In this way, it is straightforward to see that the previous two conditions ($\epsilon_H\ll3$ and $|\eta_H|\ll3$) guarantee, respectively, that the first terms of the Friedman and inflaton evolution equations:
\begin{align}
\frac{\dot\chi^2}{2}+U-3M_P^2H^2=0\,,\quad \ddot\chi+3H\dot\chi+U'=0\,.
\end{align}
are negligible, which is precisely the definition of the slow-roll approximation to the inflaton's dynamics. For more details on this approximation and the relation between the potential and the Hubble slow-roll parameters we point the reader to reference \cite{Liddle:1994dx} and the appendix of \cite{Ballesteros:2014yva}.

\subsection{Mukhanov-Sasaki formalism}
In order to compute reliably the mass and abundance of PBHs as well as the duration of inflation, it is necessary to describe accurately the dynamics of the inflaton around the (possibly deformed) inflection point of the potential, which corresponds to distance scales much smaller than those of the CMB. It turns out that the (simplest) approximation, equation \eq{app}, can fail badly in that region. The inaccuracy of \eq{app} can be easily understood by considering an exact inflection point. In that case, \eq{app} diverges when $\epsilon_U=0$, although the actual power spectrum remains finite, thanks to the non-vanishing velocity of the field. Similarly, computing the number of e-folds with the usual approximation consisting in integrating $1/\sqrt{\epsilon_U}$ over a range of field values leads to a substantial error if the potential has an (either approximate or exact) inflection point. The impossibility of using \eq{app} for describing PBH formation from a very flat inflationary potential has also been recently pointed out in \cite{Germani:2017bcs} and \cite{Motohashi:2017kbs}. 

In addition, the authors of \cite{Motohashi:2017kbs} have also shown (using several toy models) that the application of the approximation \eq{PH} --although performing better than \eq{app}-- may also lead to a wrong estimate of the mass of the PBHs. The same issue was discussed as well about ten years ago in \cite{Chongchitnan:2006wx}, which suggested --also in the context of PBH formation from inflation-- that the approximation \eq{PH} can render well the shape of the peak of the primordial power spectrum provided that the function $\epsilon_H(N_e)$ does does not grow well above unity. We will show that even if $\epsilon_H<1$, the approximation \eq{app} fails if $\eta_H$ becomes larger than 3, breaking the slow-roll approximation. 

In such a situation, an exact calculation of the primordial spectrum using the Mukhanov-Sasaki formalism \cite{Sasaki:1986hm,Mukhanov:1988jd} must be done. For the near-inflection points that we consider, this turns out to be the case, as it can be checked numerically. The Mukhanov-Sasaki equation is
\begin{align} \label{MS}
\frac{d^2 u_k}{d\tau^2}+\left(k^2-\frac{1}{z}\frac{d^2 z}{d \tau^2}\right)u_k=0\,,
\end{align}
where $\tau$ denotes conformal time ($a\,d\tau=dt$) and 
\begin{align}
u=-z\,\mathcal{R}\,,\quad z=\frac{a}{\mathcal{H}}\frac{d\phi}{d \tau}\,,
\end{align}
being $\mathcal{R}$ the comoving curvature perturbation, whose spectrum we are interested in. For scales $k \gg \mathcal{H}=a\,H$, the function $u$ is assumed, as usual, to be in the Bunch-Davies vacuum, so that its Fourier transform $u_k$ satisfies
\begin{align} \label{bd}
u_k\rightarrow\frac{e^{-i\,k\,\tau}}{\sqrt{2k}}
\end{align}
in the asymptotic past.  During inflation, $u_k$ evolves in time according to \eq{MS}. After Hubble crossing (i.e.\ at the time when $k =\mathcal{H}$), each mode $u_k$ starts approaching a constant value, which can be found solving \eq{MS} until $k$ is sufficiently smaller than $\mathcal{H}$. The primordial power spectrum for $\mathcal{R}$ can then be obtained as 
\begin{align} \label{true}
\mathcal{P}_\mathcal{R}=\frac{k^3}{2\pi^2}\left|\frac{u_k}{z}\right|^2_{k\ll \mathcal{H}}\,.
\end{align} 
The equation \eq{MS} can be written using $N_e$ as time variable: 
\begin{align} \label{MSNe}
\frac{d^2\, u_k}{d N_e^2}+(1-\epsilon_H)\frac{d\, u_k}{d N_e}+\left[\frac{k^2}{\mathcal{H}^2}+(1+\epsilon_H-\eta_H)(\eta_H-2)-\frac{d(\epsilon_H-\eta_H)}{d N_e}\right]u_k=0\,.
\end{align}
For numerical purposes it is convenient to solve separately for the real and imaginary parts of each mode $u_k$, in conjunction with the background field equation \eq{evo}.\footnote{The equation \eq{MSNe} involves the third and second derivatives of the inflaton $\chi$ with respect to $N_e$, but they can be eliminated using \eq{evo} and its first derivative.} The initial conditions for \eq{MSNe} are then set from \eq{bd} as follows:
\begin{align} \label{initc}
\mathbf{Re}(u_k)=\frac{1}{\sqrt{2k}}\,,\quad \mathbf{Im}(u_k)=0\,,\quad \mathbf{Re}\left(\frac{du_k}{dN_e}\right)=0\,,\quad \mathbf{Im}\left(\frac{du_k}{dN_e}\right)=-\frac{\sqrt{k}}{\sqrt{2}\,k_i}\,,
\end{align}
where the scale $k_i$ is chosen for each $k$ in such a way that $k_i\ll k$, so that the integration for each mode is started at a value of $N_e$ corresponding to $\mathcal{H}=k_i$. In practice, it is sufficient to choose $k_i$ a hundredth or a thousandth times smaller than the wavenumber of the mode of interest. Then, each mode has to be evolved sufficiently long, until $|u_k/z|$ reaches a constant value, were \eq{true} is evaluated. The properly normalized primordial spectrum can 
conveniently be obtained knowing its amplitude at some scale (such as $k_{0.05}$) and using that the ratio ${\mathcal{P}_\mathcal{R}(k)}/{(k^3|u_k|^2)}$ is, by construction, independent of $k$.

\section{Numerical search strategy}
\label{Numericalanalysis}

 The connection between the number of e-folds $N_e$ and the comoving distance scale $k$, is established, as we have explained, through the Hubble crossing condition $k=a\,H=\mathcal{H}$, which implies that $N_e\propto\log(k/H)$. Concretely, we write 
\begin{align} \label{kH}
k=k_*\frac{H(N_e)}{H_*}e^{N_e-N_e^*}\,,
\end{align}
where we link the fiducial scale, $k_*$ of \eq{custom}, to the time $t_*(N_e^*)$ for which $H$ takes the value $H_*$. In other words, the number of e-folds elapsed between the times at which $k=\mathcal{H}$ and $k_*=\mathcal{H_*}$  is $\Delta N_e^*=N_e-N_e^*$. In our numerical scans of the potentials \eq{V1} and \eq{V2} we set the ``initial'' value $\chi_*$ of the inflaton (that corresponds to $N_e^*$) in such a way that the scalar spectral index, $n_s$, in the slow-roll approximation, equation \eq{index}, satisfies its CMB constraint (at $k_*$) within a $\sim 3\sigma$ confidence interval. We then solve the equation \eq{evo} imposing the slow-roll attractor \eq{att} as initial condition for the velocity $d\chi/d N_e$ at $N_e^*$ and determine the number of e-folds produced until the end of inflation via the condition $\epsilon_H=1$. We also compute the tensor-to-scalar ratio, $r$, and the running of the scalar spectral index, $\alpha$, at $k_*$ --using the slow-roll expressions \eq{index} and \eq{tens}-- and we check that they remain compatible as well with the CMB constraints for a sufficient number of e-folds.  This is generically guaranteed if the variation of the first three slow-roll parameters (which are needed to obtain $n_s$, $r$ and $\alpha$) is sufficiently slow for several e-folds around $N_e^*$. We also check the validity of the slow-roll approximation at the scales probed by the CMB using the Mukhanov-Sasaki equation. Integrating the evolution equation for the inflaton beyond $\epsilon_H=1$, we get the number of e-folds that it takes the field to arrive to the minimum of its potential at $\chi=0$.
This also allows to include in the analysis cases for which there is a temporary stop of inflation (while $\epsilon_H>1$). After having obtained $\chi$ as a (numerical) function of $N_e$, we can compute the power spectrum of primordial perturbations as a function of this trajectory using \eq{PH}, and then convert the result to a function of $k$ via \eq{kH}. Then, as we already mentioned, we also solve the Mukhanov-Sasaki equation as explained in the previous section and compare the resulting primordial spectrum to the approximation \eq{PH}. \\\newline
We look for potentials that satisfy the following requirements:
\begin{enumerate}[label=\textbf{\roman*})]
\item {\it Compatibility with the current CMB constraints on the primordial spectra}.\vspace{0.2cm}\\
The 2015 Planck analysis on inflationary parameters \cite{Ade:2015lrj} indicates that the scalar spectral index at the scale $k_{0.05}= 0.05$ Mpc$^{-1}$ is about 0.965 with a (remarkably low) $\sim 0.5\%$  error at 68\% C.L. The concrete central value and error depend on the specific correlations of the channels (temperature, $T$, and $E$-mode polarization, $E$) that are considered, the datasets with which Planck data are combined (e.g.\ baryonic acoustic oscillations) and the assumptions that are made about the primordial power spectrum. For reference, in our work we take
\begin{align}\label{nsvalue}
n_s = 0.9650\pm 0.0050\quad {\rm at}\quad k_{0.05}\quad {\rm and}\quad 68 \% \quad {\rm C.L.}
\end{align} 
This value very closely agrees with the outcome of the fit allowing for a non-zero $r$, assuming $\alpha=0$ and considering $TT$, $EE$ and $TE$ correlations together with low-$\ell$ polarization data \cite{Ade:2015lrj}. As we will see, the models we consider tend to predict $r\sim 0.03$ at $k_{0.05}$, which is below the upper bound \eq{rbound}, and values of $\alpha$ well compatible with the constraint
\begin{align}
\alpha=-0.009 \pm 0.008\quad {\rm at}\quad k_{0.05}\quad {\rm and}\quad 68 \% \quad {\rm C.L.}\,,
\end{align}
which is also derived from the same data and assumptions mentioned above \cite{Ade:2015lrj}. Finally, the amplitude of the primordial perturbations, as constrained by the Planck collaboration, is
\begin{align} \label{height}
A_s=2.2\pm 0.1\times 10^{-9} \quad {\rm at}\quad k_{0.05}\quad {\rm and}\quad 68 \% \quad {\rm C.L.}\,. 
\end{align}
This condition can easily be satisfied choosing appropriately a global scaling factor in the potentials \eq{V1} and \eq{V2}; e.g. the parameter $\lambda_0$ in the second case. Indeed, what really matters of \eq{height} for our purposes is not so much its precise value, but its order of magnitude in comparison to that needed for generating PBHs, which we discuss below.

\item {\it A number of e-folds in the range $\sim$ 50 -- 60, between the time that today's largest observable scales ($k\sim k_{0.001} = 10^{-4}$ {\rm Mpc}$^{-1}$) exited the Hubble radius and the time at which inflation ended.}\vspace{0.2cm}\\
This amount of inflation is approximately what is needed to solve the horizon and flatness problems of the Universe. The precise value depends on the scale of inflation, $H$ (which has not been measured yet), and the details of the reheating process, whose theoretical computation requires knowing how the inflaton couples to other degrees of freedom. Typical models cannot produce more than {$\Delta N_e^{0.001}\sim 65$} e-folds between the time when $k_{0.001}$ becomes equal to $aH$ during inflation and the end of the process \cite{Dodelson:2003vq,Liddle:2003as}. As explained above, we determine the number of e-folds of inflation by integrating the equation \eq{evo} from an initial value of the inflaton $\chi_*$ for which the CMB constraints (at $k_*=k_{0.05}$) are met. The amount of inflation produced between $k_{0.001}$ and $k_{0.05}$ is approximately $\log(0.05/0.001)\simeq 3.9$ e-folds and can be computed exactly (again, by integrating the dynamics of the inflaton) for each model. 
Given this, we can look for models for which $\Delta N_e^*=N_e-N_*$ at the end of inflation is approximately in the range
\begin{align}\label{Nerange}
\Delta N_e^*\in\,\sim [45,55]\,. 
\end{align}
In our results we quote the value of $\Delta N_e^*$ as well as the total number of e-folds from the scale $k_{0.001}$ to the end of inflation, which we denote by $\Delta N_e^{0.001}$.

\item {\it A spike in the amplitude of scalar perturbations at a scale $k_{PBH}$ corresponding to PBH masses in an interesting window for DM, and with a height of about seven orders of magnitude more than the spectrum at CMB scales.}\vspace{0.2cm}\\
In particular, we  anticipate that the models we propose generate PBHs in the approximate mass range of 
\begin{align} \label{appM}
10^{-16.5} M_\odot\quad \text{to}\quad 10^{-13} M_\odot\,. 
\end{align} 
As shown in figure \ref{fig:omega} this is the range of masses for light PBHs that has the potential for explaining a large amount of the DM abundance. If we neglect entirely the bounds from neutron star capture in globular clusters (NS) and take the most conservative microlensing analysis of Subaru data (HSC), we see that PBHs of $\sim 10^{-13} M_\odot$ may explain the totality of the DM. The window gets enlarged down to $\sim 10^{-14} M_\odot$ if we also disregard the bound from white dwarf explosions (WD). Taking $\sim 10^{-13} M_\odot$ and using \eq{Mk}, we see that $k_{PBH}$ must be around $5\times 10^{12}$ Mpc$^{-1}$, which is many orders of magnitude smaller than any scale that can be probed with large scale structure data of the Universe.   Using \eq{Nesimp} we see that this corresponds to $\Delta N_e^*\simeq 33$.

\end{enumerate}

Satisfying simultaneously the (mutually competing) conditions that we have just enumerated cannot generically be done with a random potential with an inflection point --whether exact or approximate-- and requires a delicate balance between the parameters of the model. We have found that a shallow local minimum instead of an exact inflection point helps to fulfill them. In this situation, the classical rolling of the inflaton field can be considerably slowed down, before overcoming the local maximum --that is also inevitably generated-- and heading to the true minimum of the potential at $\phi=0$. The slow velocity of the field when it climbs out of the minimum boosts the power spectrum. This effect can be intuitively understood from equation \eq{PH}, where $\epsilon_H$ is given by \eq{eH}. Since $H$ is approximately constant around the deformed inflection point, the amplitude of the spectrum is controlled by the smallness and (as we will also see) the rate of change of $\epsilon_H$.\footnote{The regime in which the slope of the potential becomes negligible in comparison to the acceleration of the field is sometimes called ``ultra slow-roll'', see e.g.\ \cite{Tsamis:2003px,Kinney:2005vj,Martin:2012pe,Dimopoulos:2017ged}.} This kind of dynamics (using a local minimum) for producing PBHs has been noticed also in~\cite{Kannike:2017bxn}, where the approximation \eq{app} was used. However, we find that the boost in the power spectrum can only be well captured --when it is large enough to form PBHs-- by the Mukhanov-Sasaki equation. We emphasize the relevance of an accurate calculation of the spectrum, due to the exponential dependence of the mass fraction on the fluctuations; see equations \eq{eq:beta}--\eq{eq:beta3}.

In order to investigate the feasibility of PBH production from a local minimum, we have performed numerical scans of the parameter spaces of the potentials \eq{V1} and \eq{V2}, imposing the existence of such a feature. One can then look for models where the  velocity of the field around the feature is sufficiently slow to generate a large peak of scalar perturbations, but such that the field has enough slow-rolling inertia  to overcome the barrier and avoid getting trapped in the minimum. Clearly, the field may generate a significant number of e-folds while climbing up the local maximum. This amount of inflation must be such that the scale $k_{PBH}$ falls in an adequate range to generate PBHs of the appropriate mass to satisfy the bounds described above; see figure~\ref{fig:omega}. For the two kinds of potentials that we consider we have found a strong correlation between $n_s$ and the position and the height of the peak of the power spectrum; and, correspondingly, the mass and abundance of PBHs.
Namely, the models which comply with the requirements listed above tend to have $n_s$ below the currently measured central value of \eq{nsvalue}.

\section{Primordial black holes from a radiative plateau}
\label{pbhRP}

\begin{figure}[t]
\begin{center}
\includegraphics[width=0.82 \textwidth]{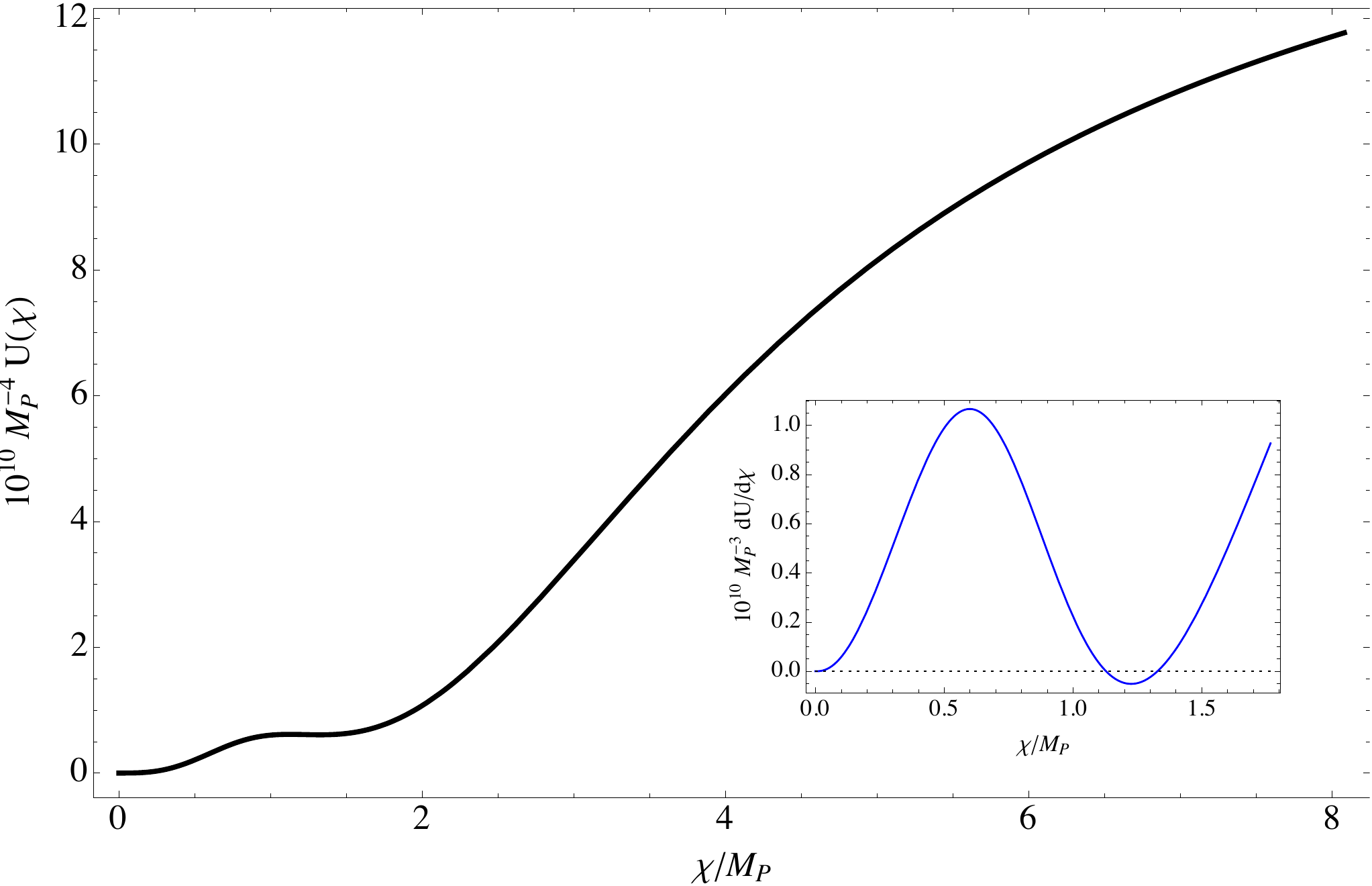} \vspace{0cm}
\caption{\em \small \label{fig:T1} {\bfseries } Inflationary potential $U(\chi)$ for the first example of Tables \ref{tab:dataparameters} and \ref{tab:data}. The inset shows the derivative of the potential between $\chi=0$ to the region around $\chi = M_P$ where PBHs are produced. The number of e-folds that occur in this region is of $\mathcal{O}(10)$. The potential realizes the qualitative properties we described with figure~\ref{fig:scheme}.}
\end{center}
\end{figure}

We will focus first on the potential \eq{V2}, which we can rewrite, neglecting higher order terms in the logarithm expansion, as
\begin{align} \label{V22}
V(\phi)=\frac{\lambda_0}{4!}\left(1-2(1+b_1)\log\frac{\phi^2}{\phi_0^2}+2(1+b_2)\left(\log\frac{\phi^2}{\phi_0^2}\right)^2\right)\phi^4\,.
\end{align}
With this notation, if $b_1=b_2=0$, the potential \eq{V22} has an exact inflection point at $\phi=\phi_0$. Hence, the parameters $b_1$ and $b_2$ characterize the level of deformation of the plateau that will lead to PBH formation. As explained in \cite{Ballesteros:2015noa}, the functional form of the potential \eq{V22} can be understood by considering a renormalizable potential in the limit of a large field $\phi$, where the corresponding quartic term dominates. The (logarithmic) effect of loop corrections to the potential can then be encoded in an effective quartic coupling $\lambda(\phi)$, leading to a potential $\lambda(\phi)\phi^4$ as in \eq{V22}. These logarithmic corrections are motivated by the fact that the inflaton needs to couple to other fields (and ultimately to the Standard Model) in order to reheat the universe after inflation \cite{Ballesteros:2015noa}.  The expression \eq{V22} thus originates from the general Coleman-Weinberg effective potential \cite{Coleman:1973jx} and can be obtained setting the renormalization scale, $\mu$, to be proportional to either $\phi_0$ or $\phi$, with both choices leading in the end to the same functional form. The second choice shows straightforwardly that the parameters $b_1$ and $b_2$ are directly related to the beta function $\beta_\lambda$ of the effective quartic coupling $\lambda$ and its first derivative at $\phi =\phi_0$ \cite{Ballesteros:2015noa}.  Concretely, 
\begin{align} \label{betas}
b_1=-1-\frac{1}{4}\frac{\beta_\lambda}{\lambda}\bigg|_{\phi=\phi_0}\,,\quad b_2=-1+\frac{1}{16\,\lambda}\frac{d\,\beta_\lambda}{d\log\mu}\bigg|_{\phi=\phi_0}\,,
\end{align}
where $\beta_\lambda =d\lambda/d\log \mu$. Therefore, an inflection point can only become manifest once two-loop effects are taken into account, since ${d\,\beta_\lambda}/d\log \mu$ is of order two in the loop expansion. 

Using the one-loop renormalization group (RG) improvement of the tree-level potential, it is possible to determine the minimal matter content that is needed to generate a potential of the form \eq{V22} with an inflection point.\footnote{This technique resums the leading logarithms at all loop orders \cite{Coleman:1973jx}, therefore accounting for the main contribution to the two-loop order term in \eq{V22}.}  It turns out that coupling a scalar $\phi$ to fermions and to another scalar (weakly coupled to the fermions) or to a $U(1)$ gauge group is sufficient \cite{Ballesteros:2015noa}. A well-known (but non-minimal) example of this is the Standard Model of particle physics, which would exhibit an inflection point at large Higgs values if the mass of this boson and the top quark were in an appropriate ratio, which is actually not too far from the actual current experimental values. 

For the purpose of analyzing the viability of \eq{V22} for generating PBHs of an adequate mass for being a substantial part of the DM, we will consider the parameters $\lambda_0$, $b_1$, $b_2$ and $\phi_0$ as constants. {They can be approximately matched to the parameters of specific particle physics models such as the ones proposed in \cite{Ballesteros:2015noa} by using the one-loop effective potential and the one-loop RG equations, together with the conditions \eq{betas}. Such an analysis shows that the perturbative behaviour of the outcoming models is well under control.

For $b_1\neq-1\neq b_2$, a potential with the qualitative features shown in figure \ref{fig:scheme} can only be obtained provided that the non-minimal coupling to gravity, $\xi$, of equation \eq{act} produces an appropriate flattening in the Einstein frame. This is indeed possible due to the running of $\xi$, whose renormalization group equation is, as it is well-known, proportional to $\xi+1/6$; vanishing in the conformal case, see \cite{Buchbinder:1992rb}. At one-loop order, we can parametrize this running with
\begin{align}
\xi(\phi)=\xi_0\left(1+b_3\log\frac{\phi^2}{\phi_0^2}\right)\,,
\end{align}
in the same fashion as for the effective quartic coupling $\lambda(\phi)$. In the large field limit, the effective potential in the Einstein frame approaches asymptotically the constant value
\begin{align}
\tilde V(\phi\rightarrow\infty) =\frac{\lambda_0}{12}\frac{1+b_2}{b_3^2\, \xi_0^2}M_P^4\,,
\end{align}
generating an inflationary plateau. Therefore, the radiative corrections to a simple renormalizable potential non-minimally coupled to gravity may in principle be sufficient to generate both primordial inflation and PBHs accounting for the DM of the Universe.

\begin{figure}[]
\begin{center}
\includegraphics[width=0.82 \textwidth]{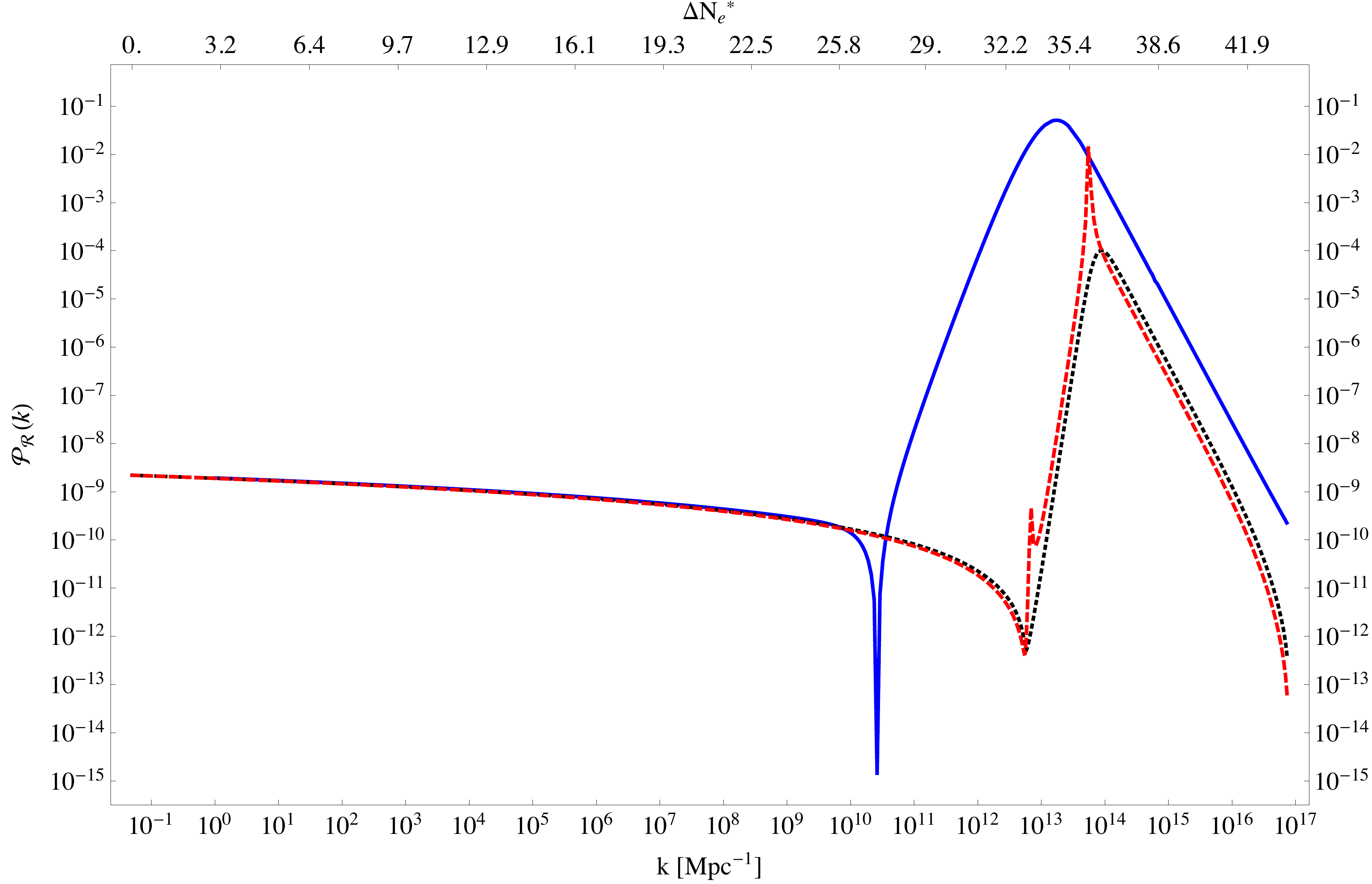} \vspace{0cm}
\caption{\em \small \label{fig:T2} {\bfseries } Blue-continuous line: primordial power spectrum as a function of the number of e-folds and the comoving wavenumber for the potential of Figure \ref{fig:T1}, computed using the Mukhanov-Sasaki formalim. The black-dotted and the red-dashed curves are the approximations \eq{PH} and \eq{app}, respectively. In this figure the number of e-folds, $\Delta N_e^*$, is set to zero at $k_*=0.05$ Mpc$^{-1}$ and the spectrum is cut approximately where inflation ends, at $\Delta N_e^*=43.5$.}
\end{center}\vspace{0.5cm}
\begin{center}
\includegraphics[width=0.82 \textwidth]{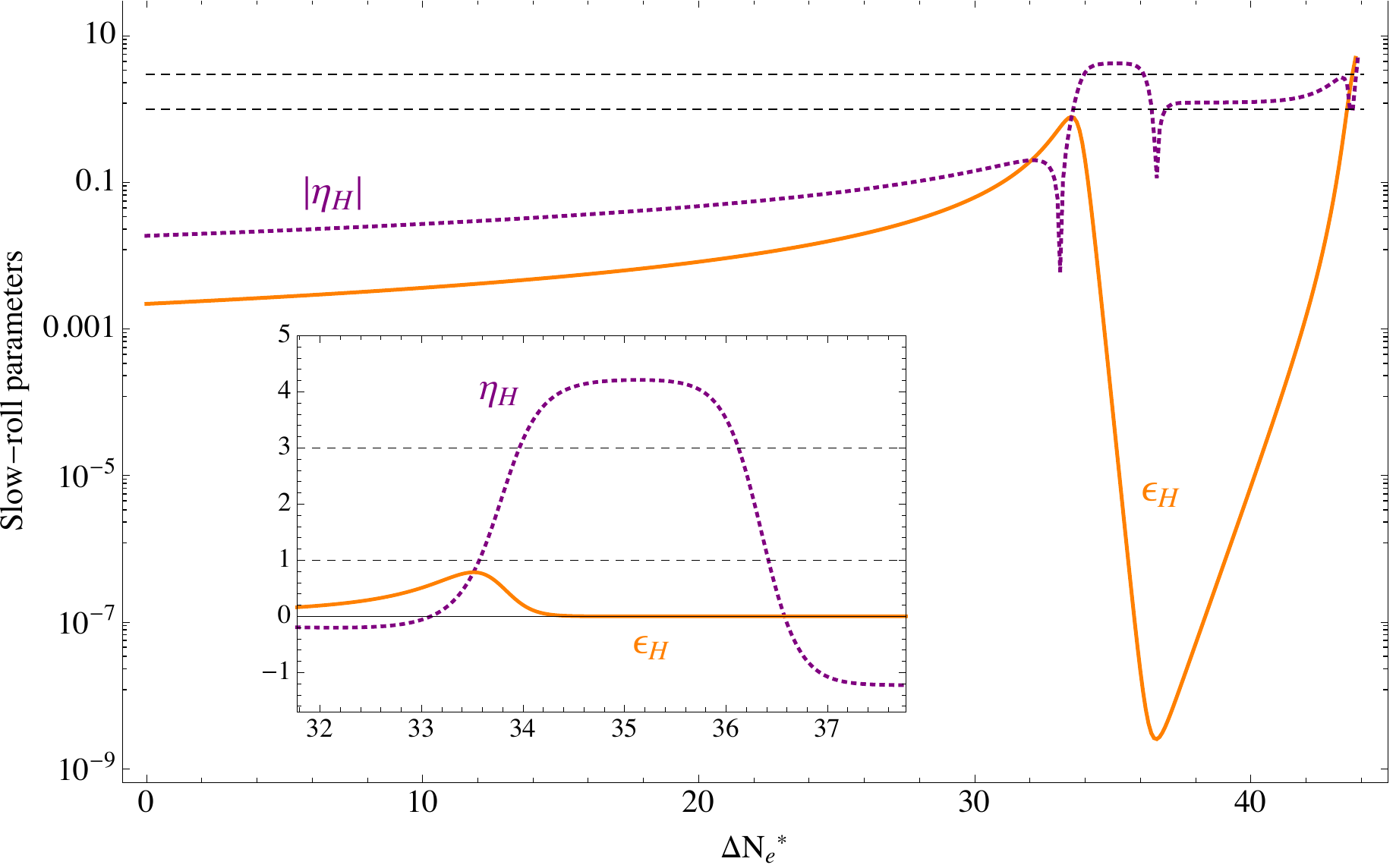} \vspace{0cm}
\caption{\em \small \label{fig:T3} {\bfseries } Slow-roll parameters $\epsilon_H$ (orange) and $\eta_H$ (purple) as functions of the number of e-folds (with the same convention as in Figure \ref{fig:T2}) for the potential of Figure \ref{fig:T1}. The horizontal dashed lines indicate the values 1 and 3. The inset zooms into the relevant range of e-folds for PBH formation, where the slow-roll approximation fails.}
\end{center}
\end{figure}

Following the strategy outlined in Section~\ref{Numericalanalysis}, we have found choices of parameters that give inflationary potentials approaching the three requirements listed in that section:\ compatibility with the CMB, enough inflation and PBHs in the low-mass window for DM. We give two examples of parameter choices in table~\ref{tab:dataparameters}. The corresponding derived cosmological parameters and the PBHs mass and abundance for these examples are given in table~\ref{tab:data}. {These examples fall into the low-mass window anticipated in \eq{appM}. In fact, all the viable examples that we have found belong in this window. In particular, we have not found any choice of parameters that gives a substantial abundance of PBHs in the mass regions constrained by EROS and the CMB (see figure \ref{fig:omega}) {\it and} that is also in agreement with the constraints on $n_s$, $r$ and the number of e-folds. Producing a large abundance of PBHs with such masses would require the inflationary plateau and the approximate inflection point (see figure \ref{fig:scheme}) to be closer to each other than in the case of the examples of tables \ref{tab:dataparameters} and \ref{tab:data}; but the separation between $\phi_*$ and $\phi_0$ cannot be arbitrarily small; see also equation~\eq{Nesimp}.} As we already mentioned in Section~\ref{Numericalanalysis}, {the examples of the tables} have low values of $n_s,$ but still within $\sim 3\sigma$ from the central value of \eq{nsvalue}. For reasons of computational feasibility we have had to use the approximation \eq{PH} as a proxy in a preliminary parameter scan which allowed us to identify potential regions of interest. Therefore, we cannot entirely exclude the possibility that viable examples with higher values of $n_s$ could be found if a more efficient search (e.g.\ with a Monte Carlo method) using directly the Mukhanov-Sasaki equation is performed. 

Figure \ref{fig:T1} shows the potential $U(\chi)$ for the first example in tables \ref{tab:dataparameters} and \ref{tab:data}. The corresponding spectrum of scalar primordial perturbations and the Hubble slow-roll parameters are displayed in figures \ref{fig:T2} and \ref{fig:T3}, on page~\pageref{fig:T2}. The potential has an asymptotic plateau, where inflation starts, and a local shallow minimum at lower field values. This feature produces a very large enhancement of the power spectrum at about 35 e-folds, when the inflaton rolls slowly climbing out of the local minimum towards the local maximum. As shown in figure \ref{fig:T2}, although the approximations \eq{PH} and \eq{app} can reproduce to some extent the qualitative behaviour of the actual spectrum computed with the Mukhanov-Sasaki equation, none of them can be used to obtain the peak height and position, which are needed to get the PBH abundance and mass. Notice that the spikes on the spectrum from \eq{app} are unphysical divergences appearing where $U'(\chi)=0$, whereas the dip in the actual solution (with the Mukhanov-Sasaki equation) at around 27 e-folds is actually a smooth feature. 

The reason for the failure of the approximation \eq{PH}, which underestimates the peak value of the spectrum, is illustrated with figure \ref{fig:T3}. The absolute value of $\eta_H$ rises above 3 around the region where PBHs form, meaning that the slow-roll approximation leading to \eq{PH} does not work, due to the second time derivative of $\chi$ becoming non-negligible. Notice also that $\epsilon_H$ remains below 1 up until the end of inflation, so that there is no temporary stop of the accelerated expansion of the Universe. However, $\epsilon_H$ does become close to 1 ($\epsilon_H\simeq 0.8$)  at around 33.4 e-folds, just before becoming nine orders of magnitude smaller. This behaviour is due to the strong slowdown of the inflaton when it rolls away from the local minimum of the potential. It is this enhanced slow-roll which produces PBHs with a mass peak given by the expression \eq{massintro} at the scale $k(N_e\simeq 35)\simeq 10^{14}$~Mpc$^{-1}$, see table~\ref{tab:data} and figure~\ref{fig:omega}. 

It is worth stressing that the PBH abundance has an exponential dependence on the critical threshold density for collapse $\delta_c$. As shown in table~\ref{tab:data}, the examples we provide can account for a significant fraction of the DM abundance for choices of  $\delta_c$ close to the values suggested in~\cite{Carr:2016drx,Musco:2004ak,Musco:2008hv,Musco:2012au,Harada:2013epa}, i.e. $\delta_c\simeq0.45$. There is still a deal of theoretical uncertainty on the actual value of $\delta_c$ that is needed for a collapse leading to PBH formation, and this can actually depend on the shape of the primordial power spectrum. We have chosen values of $\delta_c$ which lead to PBH abundances that nearly top the maxima allowed by current constraints. This is shown in figure~\ref{fig:omega}, where we have plotted the resulting (rather monochromatic) distributions for our numerical examples. 

\begin{table}[t]
 \small
  \centering
 \renewcommand\arraystretch{2.2}
\begin{tabular}{|c|c|c|c|c|c|c|}
\hline
\# & $\lambda_0$ & $\phi_0$ & $\xi_0$ & $b_1$ & $b_2$ & $b_3$ \\
\hline
\hline
1 & $6.9 \times 10^{-10}$ & 1.57353 & 0.21674 & $-1.199376$ & 0.022616 & 0.83335
\tabularnewline
\hline
\hline
2 & $2.2 \times 10^{-6}$ & 0.236027 & 6.32365 & $-0.842689$ & $-0.249409$ & 1.69240
\tabularnewline
\hline
\end{tabular} 
\caption{\it Two examples of parameter choices for the potential of Section~\ref{pbhRP}.}
\label{tab:dataparameters}

\bigskip

\small
  \centering
 \renewcommand\arraystretch{2.2}
\begin{tabular}{|c|c|c|c|c|c|c|c|c|c|c|}
\hline
\# & $\phi_*$ & $\Delta N_e^*$ & $\Delta N_e^{0.001}$ & $n_s$ & $r$ & $\alpha$ & $ \mathcal{P_R}^{peak} $ & $M_{PBHs}^{peak}/M_\odot$ &$\delta_c $& $\Omega_{PBH}/\Omega_{DM}$\\
\hline
\hline
1 & 12.30539 & 43.5 & 47.4 & 0.9531 & 0.036 &  $-0.0016$ &  0.048 & $8.3\times 10^{-15}$ & 0.51 & 0.039
\tabularnewline
\hline
\hline
2 & 1.494802 & 53.9 & 57.8 & 0.9503 & 0.027&  $-0.0018$ &  0.042 & $6.8\times 10^{-15}$& 0.47 & 0.832
\tabularnewline
\hline
\end{tabular} 
\caption{\it Results for the two examples of table~\ref{tab:dataparameters}. $\phi_*$ is the value of $\phi$ corresponding to {$k_*=0.05${Mpc}$^{-1}$,} at which the primordial parameters are given. $\Delta N_e^*$ and $\Delta N_e^{0.001}$ are the numbers of e-folds from $k_{*}$ and $k_{0.001}$, respectively, to the end of inflation. We also give the peak value of the primordial spectrum, the mass of the corresponding PBHs and their fractional abundance, computed assuming the threshold $\delta_c$. We recall that the PBH abundance is very sensitive to the value of $\delta_c$. For instance, in the first example, decreasing $\delta_c$ by two percent corresponds to a factor of five enhancement of $\Omega_{PBH}$. Indeed, barring the constraints of Figure \ref{fig:omega}, both examples allow $\Omega_{PBH}/\Omega_{DM}=1$.}
\label{tab:data}
\end{table}

As discussed in section~\ref{sec:abundance}, the PBH abundance also depends sensitively on the shape and height of $\mathcal{P_R}$ through equations \eq{eq:beta} and \eq{eq:beta1}. This, by itself, implies that some level of fine-tuning is always going to be required for PBH production with the adequate cosmological density to account for (at least a significant part of) the DM. In the context of our model, the peak of the spectrum is determined by the velocity and acceleration of the inflaton while the PBHs are generated. This translates into a strong sensitivity of the PBH abundance on the local properties of the deformed plateau, which is controlled by the parameters $b_1$ and $b_2$. In practice, one can easily check that given some value for $b_1$, the parameter $b_2$ determines the peak height. We would like to stress, though, that this sensitivity of the PBH abundance to the parameters of the Lagrangian is by no means a specific characteristic of our model, and is to be expect in any model of inflation --single-field or not-- aimed to describe PBH production, because it boils down to equations \eq{eq:beta} and \eq{eq:beta1}. 

A second (and much lesser) possible source of tuning comes (a priori) from the simple fact that among the vast range of masses in between the limits set by the Hawking radiation and the CMB, some values are much more constrained than others, see figure \ref{fig:omega}. In order for the PBHs to have masses in the low-mass window that we have discussed, the peak in the primordial spectrum has to be localized at specific scales, which means that the deformed plateau has to be reached by the inflation within a certain range of e-folds since the largest observable scales exited the Hubble radius. For instance, the plateau responsible for PBH formation cannot be too close to the asymptotic inflationary plateau that starts at larger field values, because if it were the PBHs would turn out to be too massive. Notably, both of the models we propose {generically} tend to generate PBHs in an interesting mass range once we impose sufficient e-folds to solve the horizon and flatness problems. 

We would like to conclude this section by mentioning reference  \cite{Ezquiaga:2017fvi}, which considers an action for a scalar field that can be formally obtained from ours by setting $b_1=-1$, inspired from \cite{Hamada:2014iga,Bezrukov:2014bra}. This choice amounts to eliminating the one-loop correction to the effective quartic coupling {\it at a specific scale}; see \eq{betas}. When a near-inflection point is arranged in this case (by choosing $b_3$ appropriately), the remarks that we have made concerning the invalidity of the slow-roll approximations for computing the primordial spectrum remain valid.

\section{Primordial black holes from a polynomial potential}
\label{pbhPOL}

Here we focus on the potential of Equation \eq{V1}: $V(\phi)=a_2\, \phi^2+a_3\, \phi^3+a_4\,\phi^4$. As already mentioned, a non-minimal coupling to gravity, $\xi$, is necessary in order to flatten the potential at  field values larger than the approximate inflection point, to obtain an inflationary scenario compatible with CMB observations and at the same time with the production of PBHs. It is convenient to write the Einstein frame potential $\tilde V(\phi)$ of equation \eq{potentialtilde} as follows:
\begin{align} \label{Vpol}
\tilde V(\phi)=\frac{\lambda}{4!\left(1+\xi\,{\phi^2}/{M_P^2}\right)^2}\left[ \left(3+\xi\frac{\phi_0^2}{M_P^2}\right) 2(1+c_2)\frac{\phi_0^2}{\phi^2}-8(1+c_3)\frac{\phi_0}{\phi}  +3+\xi^2\frac{\phi_0^4}{M_P^4}  \right]\,,
\end{align}
so that it has an exact inflection point at $\phi=\phi_0$ if $c_2=c_3=0$. 
With this potential, the examples we have found that produce a sufficient number of e-folds and a large PBH abundance  in the low-mass window have low values of $n_s$, more than $3 \sigma$ away from the current central value, given in equation \eq{nsvalue}. We remind that instead with the potential of equation \eq{V2} we have found models with low values of $n_s$ but still inside the $3 \sigma$ uncertainty range (and we give one in the table which is $\sim 2\sigma$ away from the central value).
Imposing the requirements from CMB observations within a $3 \sigma$ range, we have only found models with an amplitude of the power spectrum $\ll 10^{-2}.$
This means that in these cases the abundance of the resulting PBHs populations is extremely small, unless the threshold for collapse is significantly different from $\delta_c = 0.45.$ As an example, if the peak in the power spectrum of figure \ref{fig:T2} were $P_s\simeq 10^{-3}$ and $\delta_c\simeq 0.073,$ PBHs would account for $\simeq$ 8\% of the total DM abundance. Such low values of $\delta_c$ are disfavored~\cite{Carr:2016drx,Musco:2004ak,Musco:2008hv,Musco:2012au,Harada:2013epa}.

The potential \eq{Vpol} as a function of the Jordan frame field $\phi$, has the same functional form as the toy model presented in \cite{Garcia-Bellido:2017mdw}. However, the two models are fundamentally different because in our case the potential \eq{Vpol} originates from the non-minimal coupling to gravity $\sqrt{-g}\,\phi^2\,R$ and the dynamics has to be described using $U(\chi)=\tilde{V}\left(\phi(\chi)\right)$ with the Einstein frame field $\chi(\phi)$, as we explained in Section \ref{setup}. In reference \cite{Garcia-Bellido:2017mdw} it is instead assumed that $\phi$ (and not $\chi$) is the Einstein frame inflaton. In other words: the change of frame needed to explain the ratio of polynomials in \eq{Vpol} is not implemented in the analysis of \cite{Garcia-Bellido:2017mdw}. Notice that the field redefinition from $\phi$ to $\chi$ does change the shape of the actual inflationary potential and therefore the {model} of \cite{Garcia-Bellido:2017mdw} cannot be considered a proxy to the dynamics of \eq{V1} with non-minimal coupling to $R$.  Our work also {goes beyond} \cite{Garcia-Bellido:2017mdw} because whereas we apply the Mukhanov-Sasaki formalism to compute the primordial spectrum, reference \cite{Garcia-Bellido:2017mdw}  uses the approximation \eq{PH}. {As we have discussed, this approximation may fail to reproduce the actual spectrum, depending on the behaviour of the slow-roll parameters. However, we have checked that both \eq{PH} and \eq{app} reproduce sufficiently well the peak height of the spectrum in the specific numerical example of the toy model of \cite{Garcia-Bellido:2017mdw}. In that case, $\epsilon_H<0.1$ and $|\eta_H|$ remains just below 3 around the near-inflection point. This leads to a lower and broader peak than in our models, with $\mathcal{P_R}(10^{14} \text{ Mpc}^{-1})\simeq 8\times 10^{-5}$. A value of the threshold for collapse $\delta_c$ about an order of magnitude lower than the preferred one --which is $\delta_c \simeq 0.45$~\cite{Musco:2004ak,Musco:2008hv,Musco:2012au,Harada:2013epa}-- would be needed to produced a large population of PBHs from such a peak.}  

It is also worth mentioning reference \cite{Kannike:2017bxn}, which studies PBH formation from a set-up similar to the one that we have discussed in this section. However, the quartic self-coupling of the inflation, $\lambda$, in \cite{Kannike:2017bxn} contains a field-dependent piece of the form $\theta(\chi-M)\log(\chi/M)$, where $M$ is a constant that is meant to represent the effect of a cubic coupling between the inflaton and another scalar. The motivation in  \cite{Kannike:2017bxn} to introduce this term is the claim of the authors that it helps to fit the CMB data. They compute the primordial spectrum separately above and below the threshold $\chi = M$, using the approximation \eq{PH}. However, this extra term introduces a violation of slow-roll (close to the critical region for PBH formation) that cannot be described with that approximation. Besides, the introduction of such an a abrupt change in $\lambda$ to describe a coupling between the inflaton and another field may be questioned. In the presence of such a coupling one can instead compute the full Coleman-Weinberg effective potential, and if the effect of the second field is indeed very important, the dynamics of the system is likely to be better described as a two-field model.

\section{Discussion and conclusions}
\label{sec:conclusions}

In this work we have investigated the possibility that PBHs may constitute a substantial part of the DM of the Universe. In order to take seriously this idea, a mechanism operating before the time of nucleosynthesis that is able to generate a large abundance of PBHs with adequate masses is required. In absence of a concrete and consistent mechanism for PBH formation, the idea of such objects constituting a large fraction of the DM could be argued to lack a well-grounded basis. This has been our main motivation in this paper to look for such a mechanism. PBHs can form when the comoving wavenumber of sufficiently large density fluctuations becomes comparable to the conformal Hubble scale after inflation. Within this general framework, the simplest possibility is that the inflaton itself generates primordial fluctuations that act as seeds for PBHs. We have taken on the challenge of finding a single-field model of inflation with the appropriate properties and that is well-motivated from the point of view of particle physics. Much of the difficulty to achieve this comes from the fact that the primordial power spectrum has to be enhanced at specific small distances ($k\sim 10^{14}$~Mpc$^{-1}$ for the low-mass window) by roughly seven orders of magnitude with respect to CMB scales. In what follows we summarize our findings, commenting on the main features of our work and mentioning a couple of directions in which it may be extended.
\begin{itemize}
\item Models.

The two key features of the models we have considered are a potential with an approximate plateau and a generic $\sqrt{-g}\,\phi^2\,R$ coupling between the inflaton and the Ricci scalar.

The first model we have proposed consists in a simple quartic potential that develops a near-plateau region due to radiative corrections, equation~\eq{V2}. This type of potential arises in the context of renormalizable and perturbative particle physics models \cite{Ballesteros:2015noa}. The couplings of the inflaton to other fields that correct the tree-level potential are motivated by the requirement of reheating the Universe after inflation. 

The second model that we have studied is, again, a renormalizable potential, equation \eq{V1}. In this case, differently from the previous one, the cubic and quadratic terms become comparable to the quartic term before inflation ends, generating the feature that leads to PBH formation. 

In both cases, the non-minimal coupling to $R$ softens the approximately quartic growth of the potentials at large field values, reducing the amount of primordial GW at CMB scales and thus ensuring compatibility with current constraints on the tensor-to-scalar ratio. Whereas the non-minimal coupling parameter can be assumed to be a constant in the second model, it has to have a non-negligible running (which is anyway generic) in the case of the first potential. It is worth mentioning that the values of the non-minimal coupling that lead to interesting PBH masses are of order 1 or smaller, which aids to ensure that perturbative unitarity is preserved up to very high energies, see \cite{Barbon:2009ya,Burgess:2009ea,Burgess:2010zq,Ballesteros:2016xej}.

\item Results.

We have demonstrated that the first of the models we propose, based on the potential \eq{V2}, allows successful inflation and a population of PBHs which can account for a significant fraction of the DM. The resulting PBH mass spectrum is almost monochromatic, notably, peaking around $10^{-16}-10^{-15} M_{\odot}$, in the neighbourhood of a mass range where some of the current constraints on PBHs are arguably less robust and PBHs might be all the dark matter of the Universe. Both models exhibit a tendency to produce low values of the scalar spectral index at CMB scales, $n_s$, with the highest values typically corresponding to lower PBH masses. In particular, in the second model requiring a sufficient population of PHBs to account for a large DM fraction in the low-mass window seems to imply values of $n_s$ which differ by more than $3\sigma$ with respect to the currently measured central value \eq{nsvalue}.

Investigating how to reduce this discrepancy with variants of our models is worth of further study. In this respect, let us mention that the CMB constraints that we have quoted so far on $n_s$ change, shifting its central value about $\lesssim 1 \sigma$, once the sum of the neutrino masses is included as another cosmological parameter of the base cosmological model \cite{Gerbino:2016sgw}.
The parameter scan that we have performed is not exhaustive and therefore we cannot exclude with absolute certainty that the models may hide examples contradicting the general trend on $n_s$ that we have observed. Interestingly, both models give values of the tensor-to-scalar ratio around $0.03$, within the planned reach of future experiments such as CMB-S4 \cite{Abazajian:2016yjj} and LiteBird \cite{Matsumura:2013aja}. The values of the running of the spectral index in our examples might also be tested with the next generation of probes. 

\item Slow-roll.

We have shown that an accurate computation of the primordial spectrum leading to PBH formation requires the use of the Mukhanov-Sasaki formalism. The application of the standard approximations based on the potential \eq{app} and Hubble \eq{PH} slow-roll parameters can lead to severely wrong estimates of the PBH mass and abundance (by several orders of magnitude). Concretely, the best of these two approximations, equation \eq{PH}, fails because it neglects the effect of the acceleration of the inflaton and its variation, which are encoded in $\eta_H$ and the derivative of $\epsilon_H-\eta_H$ in equation \eq{MSNe}. This conclusion applies as well to other models of inflation leading to PBH formation from a very slow phase of rolling, as it is also pointed out in \cite{Motohashi:2017kbs}. 

\item Parameter sensitivity. 

The abundance of PBHs depends critically on the threshold for collapse and the primordial spectrum peak height. The threshold for collapse is not known precisely and we have taken values close to $\delta_c=0.45$, as suggested by recent analyses~\cite{Carr:2016drx,Musco:2004ak,Musco:2008hv,Musco:2012au,Harada:2013epa}. In our models, the height of the primordial spectrum is determined by the level of deformation of the plateau that is responsible for PBH generation. Due to this, the PBH abundance at the time of formation is strongly sensitive to the parameters that control the depth of the (shallow) local minimum of the potential.\footnote{Notice that in our examples we can approximately compensate a change in $\delta_c$ with a modification of the parameters of the potential, keeping the same PBH abundance.} This sensitivity is also necessarily present in any other inflationary implementation of PBH formation. {Indeed, the generic mechanism for PBH formation summarized in Section \ref{sec:abundance} (of large fluctuations collapsing upon Hubble crossing) implies that the PBH abundance in any model of inflation aiming to implement such process will have a strong parameter dependence.} Remarkably, in our case the the PBHs masses turn to be in an interesting region for explaining DM --the low-mass window-- once a sufficient amount of inflation for solving the horizon and flatness problems is imposed. 

\item Current bounds on PBH as DM.

Although a detailed discussion is outside the scope of our work, we would like to stress the importance of further observational and theoretical studies to clarify which are the possible windows where, given the current and possible future data, PBHs may be relevant as a DM candidate. At the PBH low-mass end (limited by Hawking evaporation) the available constraints on the literature come from: Subaru microlensing~\cite{Niikura:2017zjd}, hypothetical encounters between PBHs and neutron stars in globular clusters\cite{Capela:2013yf} or white dwarfs \cite{Graham:2015apa} and lensing of gamma-ray bursts by PBHs \cite{Barnacka:2012bm}. Some of these constraints are subject to significant uncertainties. As we mentioned in the Introduction, taking the most conservative limits from Subaru and neglecting entirely the bounds from neutron star capture, the totally of the DM may be accounted for PBHs with masses in the range that the models we propose are able to produce.

\end{itemize}

Let us finally mention two directions along which it would be interesting to extend our work:

\begin{itemize}

\item GW background from PBH formation and merging. 

We can also use the Mukhanov-Sasaki formalism to obtain the primordial tensor spectrum at linear order in perturbations, solving an equation similar to \eq{MSNe} with initial conditions identical to \eq{initc}. {This shows a suppression of the tensor spectrum at the scales of PBH formation with respect to the slow-roll approximations.} More importantly, the large scalar fluctuations responsible for PBH formation source a sizable GW spectrum from second order perturbations, which may be potentially observable with laser interferometers, see e.g.\ \cite{Alabidi:2012ex,Garcia-Bellido:2017aan}. Moreover, extra GWs are generated if PBHs encounter and merge, leading to heavier BHs and thus changing their mass distribution.The spectrum of GW from these last two effects may lead to interesting constraints on our mechanism for PBH formation. 

\item Quantum fluctuations in quasi-De Sitter space.

Our analysis for the dynamics of inflation has been focused on the classical trajectory of the field, derived from the action \eq{act2}. One might wonder whether quantum fluctuations induced during inflation\footnote{We do not refer here to the quantum corrections that shape the potential \ref{V2}, but to those due to the field being in a near-de Sitter space during inflation.} may play a significant role in the dynamics,\footnote{See e.g.\ \cite{Pattison:2017mbe}, although that formalism is not directly applicable in our case.} especially in the vicinity of the local minimum that generates the PBHs, since in this region the field undergoes a phase of strong deceleration. Quite generically, the quantum fluctuations of the inflaton are of the order of $H/2\pi$ and they do not  significantly alter the classical trajectory provided that $\dot\chi/H\gg {H}/{(2\pi)}$.
This condition is equivalent to $\mathcal{P_R} \ll 1,$ which is always satisfied provided that PBHs are not overproduced, as required for consistency. Nevertheless, it may be worth exploring in some detail the role of these quantum fluctuations during inflation around the shallow minimum, since they may have interesting effects like, for instance, broadening the mass function of PBHs. 

\end{itemize}

\subsubsection*{Acknowledgments}

We thank P.\ Serpico for useful discussions {and comments on the paper.} G.B.\ thanks as well J.R.~Espinosa, R.\ Flauger, C.\ Tamarit and A.\ Westphal for useful conversations.}
The work of G.B.\ has been funded by the European Union's Horizon 2020 research and innovation programme under the Marie Sk\l{}odowska-Curie grant agreement number 656794. The work of M.T.\ is supported by the Spanish Research Agency (Agencia Estatal de Investigaci\'on) through the grant IFT Centro de Excelencia Severo Ochoa SEV-2016-0597 and the projects FPA2015-65929-P and Consolider MultiDark CSD2009-00064. G.B.\ thanks the CERN Theoretical Physics department for hospitality while part of this research was done. M.T. acknowledges the hospitality of the Instituto de  F\'isica Corpuscular in Valencia, where part of this work was done.

\bibliography{pbhBIBx_V5}

\begin{thebibliography}{10}

\bibitem{Abbott:2016blz}
B.~P. Abbott et~al.
\newblock {Observation of Gravitational Waves from a Binary Black Hole Merger}.
\newblock {\em Phys. Rev. Lett.}, 116(6):061102, 2016, 1602.03837.

\bibitem{Abbott:2016nmj}
B.~P. Abbott et~al.
\newblock {GW151226: Observation of Gravitational Waves from a 22-Solar-Mass
  Binary Black Hole Coalescence}.
\newblock {\em Phys. Rev. Lett.}, 116(24):241103, 2016, 1606.04855.

\bibitem{Abbott:2017vtc}
Benjamin~P. Abbott et~al.
\newblock {GW170104: Observation of a 50-Solar-Mass Binary Black Hole
  Coalescence at Redshift 0.2}.
\newblock {\em Phys. Rev. Lett.}, 118(22):221101, 2017, 1706.01812.

\bibitem{TheLIGOScientific:2016htt}
B.~P. Abbott et~al.
\newblock {Astrophysical Implications of the Binary Black-Hole Merger
  GW150914}.
\newblock {\em Astrophys. J.}, 818(2):L22, 2016, 1602.03846.

\bibitem{Mandel:2015qlu}
Ilya Mandel and Selma~E. de~Mink.
\newblock {Merging binary black holes formed through chemically homogeneous
  evolution in short-period stellar binaries}.
\newblock {\em Mon. Not. Roy. Astron. Soc.}, 458(3):2634--2647, 2016,
  1601.00007.

\bibitem{Belczynski:2016obo}
Krzysztof Belczynski, Daniel~E. Holz, Tomasz Bulik, and Richard O'Shaughnessy.
\newblock {The first gravitational-wave source from the isolated evolution of
  two 40-100 Msun stars}.
\newblock {\em Nature}, 534:512, 2016, 1602.04531.

\bibitem{Rodriguez:2016kxx}
Carl~L. Rodriguez, Sourav Chatterjee, and Frederic~A. Rasio.
\newblock {Binary Black Hole Mergers from Globular Clusters: Masses, Merger
  Rates, and the Impact of Stellar Evolution}.
\newblock {\em Phys. Rev.}, D93(8):084029, 2016, 1602.02444.

\bibitem{Marchant:2016wow}
Pablo Marchant, Norbert Langer, Philipp Podsiadlowski, Thomas~M. Tauris, and
  Takashi~J. Moriya.
\newblock {A new route towards merging massive black holes}.
\newblock {\em Astron. Astrophys.}, 588:A50, 2016, 1601.03718.

\bibitem{Elbert:2017sbr}
Oliver~D. Elbert, James~S. Bullock, and Manoj Kaplinghat.
\newblock {Counting Black Holes: The Cosmic Stellar Remnant Population and
  Implications for LIGO}.
\newblock 2017, 1703.02551.

\bibitem{1966AZh43758Z}
Y.~B. Zeldovich and I.~D. Novikov.
\newblock {The Hypothesis of Cores Retarded during Expansion and the Hot
  Cosmological Model}.
\newblock {\em Astronomicheskii Zhurnal}, 43:758, 1966.

\bibitem{Hawking:1971ei}
Stephen Hawking.
\newblock {Gravitationally collapsed objects of very low mass}.
\newblock {\em Mon. Not. Roy. Astron. Soc.}, 152:75, 1971.

\bibitem{Carr:1974nx}
Bernard~J. Carr and S.~W. Hawking.
\newblock {Black holes in the early Universe}.
\newblock {\em Mon. Not. Roy. Astron. Soc.}, 168:399--415, 1974.

\bibitem{Bird:2016dcv}
Simeon Bird, Ilias Cholis, Julian~B. Mu\~noz, Yacine Ali-Haimoud, Marc
  Kamionkowski, Ely~D. Kovetz, Alvise Raccanelli, and Adam~G. Riess.
\newblock {Did LIGO detect dark matter?}
\newblock {\em Phys. Rev. Lett.}, 116(20):201301, 2016, 1603.00464.

\bibitem{Clesse:2016vqa}
Sebastien Clesse and Juan Garc\'ia-Bellido.
\newblock {The clustering of massive Primordial Black Holes as Dark Matter:
  measuring their mass distribution with Advanced LIGO}.
\newblock {\em Phys. Dark Univ.}, 15:142--147, 2017, 1603.05234.

\bibitem{Sasaki:2016jop}
Misao Sasaki, Teruaki Suyama, Takahiro Tanaka, and Shuichiro Yokoyama.
\newblock {Primordial Black Hole Scenario for the Gravitational-Wave Event
  GW150914}.
\newblock {\em Phys. Rev. Lett.}, 117(6):061101, 2016, 1603.08338.

\bibitem{Carr:2016drx}
Bernard Carr, Florian Kuhnel, and Marit Sandstad.
\newblock {Primordial Black Holes as Dark Matter}.
\newblock {\em Phys. Rev.}, D94(8):083504, 2016, 1607.06077.

\bibitem{Carr:2009jm}
B.~J. Carr, Kazunori Kohri, Yuuiti Sendouda, and Jun'ichi Yokoyama.
\newblock {New cosmological constraints on primordial black holes}.
\newblock {\em Phys. Rev.}, D81:104019, 2010, 0912.5297.

\bibitem{Barnacka:2012bm}
A.~Barnacka, J.~F. Glicenstein, and R.~Moderski.
\newblock {New constraints on primordial black holes abundance from
  femtolensing of gamma-ray bursts}.
\newblock {\em Phys. Rev.}, D86:043001, 2012, 1204.2056.

\bibitem{Graham:2015apa}
Peter~W. Graham, Surjeet Rajendran, and Jaime Varela.
\newblock {Dark Matter Triggers of Supernovae}.
\newblock {\em Phys. Rev.}, D92(6):063007, 2015, 1505.04444.

\bibitem{Capela:2013yf}
Fabio Capela, Maxim Pshirkov, and Peter Tinyakov.
\newblock {Constraints on primordial black holes as dark matter candidates from
  capture by neutron stars}.
\newblock {\em Phys. Rev.}, D87(12):123524, 2013, 1301.4984.

\bibitem{Niikura:2017zjd}
Hiroko Niikura, Masahiro Takada, Naoki Yasuda, Robert~H. Lupton, Takahiro Sumi,
  Surhud More, Anupreeta More, Masamune Oguri, and Masashi Chiba.
\newblock {Microlensing constraints on $10^{-10}M_\odot$-scale primordial black
  holes from high-cadence observation of M31 with Hyper Suprime-Cam}.
\newblock 2017, 1701.02151.

\bibitem{Tisserand:2006zx}
P.~Tisserand et~al.
\newblock {Limits on the Macho Content of the Galactic Halo from the EROS-2
  Survey of the Magellanic Clouds}.
\newblock {\em Astron. Astrophys.}, 469:387--404, 2007, astro-ph/0607207.

\bibitem{Monroy-Rodriguez:2014ula}
Miguel~A. Monroy-Rodr\'iguez and Christine Allen.
\newblock {The end of the MACHO era- revisited: new limits on MACHO masses from
  halo wide binaries}.
\newblock {\em Astrophys. J.}, 790(2):159, 2014, 1406.5169.

\bibitem{Brandt:2016aco}
Timothy~D. Brandt.
\newblock {Constraints on MACHO Dark Matter from Compact Stellar Systems in
  Ultra-Faint Dwarf Galaxies}.
\newblock {\em Astrophys. J.}, 824(2):L31, 2016, 1605.03665.

\bibitem{Koushiappas:2017chw}
Savvas~M. Koushiappas and Abraham Loeb.
\newblock {Dynamics of dwarf galaxies disfavor stellar-mass black hole dark
  matter}.
\newblock 2017, 1704.01668.

\bibitem{Ali-Haimoud:2016mbv}
Yacine Ali-Haimoud and Marc Kamionkowski.
\newblock {Cosmic microwave background limits on accreting primordial black
  holes}.
\newblock {\em Phys. Rev.}, D95(4):043534, 2017, 1612.05644.

\bibitem{Poulin:2017bwe}
Vivian Poulin, Pasquale~D. Serpico, Francesca Calore, Sebastien Clesse, and
  Kazunori Kohri.
\newblock {Squeezing spherical cows: CMB bounds on disk-accreting massive
  Primordial Black Holes}.
\newblock 2017, 1707.04206.

\bibitem{Gaggero:2016dpq}
Daniele Gaggero, Gianfranco Bertone, Francesca Calore, Riley M.~T. Connors,
  Mark Lovell, Sera Markoff, and Emma Storm.
\newblock {Searching for Primordial Black Holes in the radio and X-ray sky}.
\newblock {\em Phys. Rev. Lett.}, 118(24):241101, 2017, 1612.00457.

\bibitem{Carr:1975qj}
Bernard~J. Carr.
\newblock {The Primordial black hole mass spectrum}.
\newblock {\em Astrophys. J.}, 201:1--19, 1975.

\bibitem{Carr:2017jsz}
Bernard Carr, Martti Raidal, Tommi Tenkanen, Ville Vaskonen, and Hardi Veermae.
\newblock {Primordial black hole constraints for extended mass functions}.
\newblock 2017, 1705.05567.

\bibitem{Hawking:1974rv}
S.~W. Hawking.
\newblock {Black hole explosions}.
\newblock {\em Nature}, 248:30--31, 1974.

\bibitem{Ricotti:2007au}
Massimo Ricotti, Jeremiah~P. Ostriker, and Katherine~J. Mack.
\newblock {Effect of Primordial Black Holes on the Cosmic Microwave Background
  and Cosmological Parameter Estimates}.
\newblock {\em Astrophys. J.}, 680:829, 2008, 0709.0524.

\bibitem{Blum:2016cjs}
Daniel Aloni, Kfir Blum, and Raphael Flauger.
\newblock {Cosmic microwave background constraints on primordial black hole
  dark matter}.
\newblock {\em JCAP}, 1705(05):017, 2017, 1612.06811.

\bibitem{Mediavilla:2017bok}
E.~Mediavilla, J.~Jim\'enez-Vicente, J.~A. Mu\~noz, H.~Vives-Arias, and
  J.~Calder\'on-Infante.
\newblock {Limits on the Mass and Abundance of Primordial Black Holes from
  Quasar Gravitational Microlensing}.
\newblock {\em Astrophys. J.}, 836(2):L18, 2017, 1702.00947.

\bibitem{Ivanov:1994pa}
P.~Ivanov, P.~Naselsky, and I.~Novikov.
\newblock {Inflation and primordial black holes as dark matter}.
\newblock {\em Phys. Rev.}, D50:7173--7178, 1994.

\bibitem{1990NuPhB.335..197H}
H.~M. {Hodges}, G.~R. {Blumenthal}, L.~A. {Kofman}, and J.~R. {Primack}.
\newblock {Nonstandard primordial fluctuations from a polynomial inflation
  potential}.
\newblock {\em Nuclear Physics B}, 335:197--220, April 1990.

\bibitem{Destri:2007pv}
C.~Destri, Hector~J. de~Vega, and N.~G. Sanchez.
\newblock {MCMC analysis of WMAP3 and SDSS data points to broken symmetry
  inflaton potentials and provides a lower bound on the tensor to scalar
  ratio}.
\newblock {\em Phys. Rev.}, D77:043509, 2008, astro-ph/0703417.

\bibitem{Ballesteros:2015noa}
Guillermo Ballesteros and Carlos Tamarit.
\newblock {Radiative plateau inflation}.
\newblock {\em JHEP}, 02:153, 2016, 1510.05669.

\bibitem{Array:2015xqh}
P.~A.~R. Ade et~al.
\newblock {Improved Constraints on Cosmology and Foregrounds from BICEP2 and
  Keck Array Cosmic Microwave Background Data with Inclusion of 95 GHz Band}.
\newblock {\em Phys. Rev. Lett.}, 116:031302, 2016, 1510.09217.

\bibitem{Spokoiny:1984bd}
B.~L. Spokoiny.
\newblock {Inflation and generation of perturbations in broken symmetric theory
  of gravity}.
\newblock {\em Phys. Lett.}, 147B:39--43, 1984.

\bibitem{Garcia-Bellido:2017mdw}
Juan Garcia-Bellido and Ester Ruiz~Morales.
\newblock {Primordial black holes from single field models of inflation}.
\newblock 2017, 1702.03901v5.

\bibitem{Ezquiaga:2017fvi}
Jose~Maria Ezquiaga, Juan Garcia-Bellido, and Ester Ruiz~Morales.
\newblock {Primordial Black Hole production in Critical Higgs Inflation}.
\newblock 2017, 1705.04861v2.

\bibitem{Sasaki:1986hm}
Misao Sasaki.
\newblock {Large Scale Quantum Fluctuations in the Inflationary Universe}.
\newblock {\em Prog. Theor. Phys.}, 76:1036, 1986.

\bibitem{Mukhanov:1988jd}
Viatcheslav~F. Mukhanov.
\newblock {Quantum Theory of Gauge Invariant Cosmological Perturbations}.
\newblock {\em Sov. Phys. JETP}, 67:1297--1302, 1988.
\newblock [Zh. Eksp. Teor. Fiz.94N7,1(1988)].

\bibitem{Press:1973iz}
William~H. Press and Paul Schechter.
\newblock {Formation of galaxies and clusters of galaxies by selfsimilar
  gravitational condensation}.
\newblock {\em Astrophys. J.}, 187:425--438, 1974.

\bibitem{Young:2014ana}
Sam Young, Christian~T. Byrnes, and Misao Sasaki.
\newblock {Calculating the mass fraction of primordial black holes}.
\newblock {\em JCAP}, 1407:045, 2014, 1405.7023.

\bibitem{Liddle:2000cg}
Andrew~R. Liddle and D.~H. Lyth.
\newblock {\em {Cosmological inflation and large scale structure}}.
\newblock 2000.

\bibitem{Inomata:2017okj}
Keisuke Inomata, Masahiro Kawasaki, Kyohei Mukaida, Yuichiro Tada, and
  Tsutomu~T. Yanagida.
\newblock {Inflationary Primordial Black Holes as All Dark Matter}.
\newblock {\em Phys. Rev.}, D96(4):043504, 2017, 1701.02544.

\bibitem{Ade:2015xua}
P.~A.~R. Ade et~al.
\newblock {Planck 2015 results. XIII. Cosmological parameters}.
\newblock {\em Astron. Astrophys.}, 594:A13, 2016, 1502.01589.

\bibitem{Musco:2004ak}
Ilia Musco, John~C. Miller, and Luciano Rezzolla.
\newblock {Computations of primordial black hole formation}.
\newblock {\em Class. Quant. Grav.}, 22:1405--1424, 2005, gr-qc/0412063.

\bibitem{Musco:2008hv}
Ilia Musco, John~C. Miller, and Alexander~G. Polnarev.
\newblock {Primordial black hole formation in the radiative era: Investigation
  of the critical nature of the collapse}.
\newblock {\em Class. Quant. Grav.}, 26:235001, 2009, 0811.1452.

\bibitem{Musco:2012au}
Ilia Musco and John~C. Miller.
\newblock {Primordial black hole formation in the early universe: critical
  behaviour and self-similarity}.
\newblock {\em Class. Quant. Grav.}, 30:145009, 2013, 1201.2379.

\bibitem{Harada:2013epa}
Tomohiro Harada, Chul-Moon Yoo, and Kazunori Kohri.
\newblock {Threshold of primordial black hole formation}.
\newblock {\em Phys. Rev.}, D88(8):084051, 2013, 1309.4201.
\newblock [Erratum: Phys. Rev.D89,no.2,029903(2014)].

\bibitem{Motohashi:2017kbs}
Hayato Motohashi and Wayne Hu.
\newblock {Primordial Black Holes and Slow-Roll Violation}.
\newblock {\em Phys. Rev.}, D96(6):063503, 2017, 1706.06784.

\bibitem{Ballesteros:2014yva}
Guillermo Ballesteros and J.~Alberto Casas.
\newblock {Large tensor-to-scalar ratio and running of the scalar spectral
  index with Instep Inflation}.
\newblock {\em Phys. Rev.}, D91:043502, 2015, 1406.3342.

\bibitem{Stewart:1993bc}
Ewan~D. Stewart and David~H. Lyth.
\newblock {A More accurate analytic calculation of the spectrum of cosmological
  perturbations produced during inflation}.
\newblock {\em Phys. Lett.}, B302:171--175, 1993, gr-qc/9302019.

\bibitem{Liddle:1994dx}
Andrew~R. Liddle, Paul Parsons, and John~D. Barrow.
\newblock {Formalizing the slow roll approximation in inflation}.
\newblock {\em Phys. Rev.}, D50:7222--7232, 1994, astro-ph/9408015.

\bibitem{Germani:2017bcs}
Cristiano Germani and Tomislav Prokopec.
\newblock {On primordial black holes from an inflection point}.
\newblock 2017, 1706.04226.

\bibitem{Chongchitnan:2006wx}
Sirichai Chongchitnan and George Efstathiou.
\newblock {Accuracy of slow-roll formulae for inflationary perturbations:
  implications for primordial black hole formation}.
\newblock {\em JCAP}, 0701:011, 2007, astro-ph/0611818.

\bibitem{Ade:2015lrj}
P.~A.~R. Ade et~al.
\newblock {Planck 2015 results. XX. Constraints on inflation}.
\newblock {\em Astron. Astrophys.}, 594:A20, 2016, 1502.02114.

\bibitem{Dodelson:2003vq}
Scott Dodelson and Lam Hui.
\newblock {A Horizon ratio bound for inflationary fluctuations}.
\newblock {\em Phys. Rev. Lett.}, 91:131301, 2003, astro-ph/0305113.

\bibitem{Liddle:2003as}
Andrew~R Liddle and Samuel~M Leach.
\newblock {How long before the end of inflation were observable perturbations
  produced?}
\newblock {\em Phys. Rev.}, D68:103503, 2003, astro-ph/0305263.

\bibitem{Tsamis:2003px}
N.~C. Tsamis and Richard~P. Woodard.
\newblock {Improved estimates of cosmological perturbations}.
\newblock {\em Phys. Rev.}, D69:084005, 2004, astro-ph/0307463.

\bibitem{Kinney:2005vj}
William~H. Kinney.
\newblock {Horizon crossing and inflation with large eta}.
\newblock {\em Phys. Rev.}, D72:023515, 2005, gr-qc/0503017.

\bibitem{Martin:2012pe}
Jerome Martin, Hayato Motohashi, and Teruaki Suyama.
\newblock {Ultra Slow-Roll Inflation and the non-Gaussianity Consistency
  Relation}.
\newblock {\em Phys. Rev.}, D87(2):023514, 2013, 1211.0083.

\bibitem{Dimopoulos:2017ged}
Konstantinos Dimopoulos.
\newblock {Ultra slow-roll inflation demystified}.
\newblock {\em Phys. Lett.}, B775:262--265, 2017, 1707.05644.

\bibitem{Kannike:2017bxn}
Kristjan Kannike, Luca Marzola, Martti Raidal, and Hardi Veermae.
\newblock {Single Field Double Inflation and Primordial Black Holes}.
\newblock 2017, 1705.06225.

\bibitem{Coleman:1973jx}
Sidney~R. Coleman and Erick~J. Weinberg.
\newblock {Radiative Corrections as the Origin of Spontaneous Symmetry
  Breaking}.
\newblock {\em Phys. Rev.}, D7:1888--1910, 1973.

\bibitem{Buchbinder:1992rb}
I.~L. Buchbinder, S.~D. Odintsov, and I.~L. Shapiro.
\newblock {\em {Effective action in quantum gravity}}.
\newblock 1992.

\bibitem{Hamada:2014iga}
Yuta Hamada, Hikaru Kawai, Kin-ya Oda, and Seong~Chan Park.
\newblock {Higgs Inflation is Still Alive after the Results from BICEP2}.
\newblock {\em Phys. Rev. Lett.}, 112(24):241301, 2014, 1403.5043.

\bibitem{Bezrukov:2014bra}
Fedor Bezrukov and Mikhail Shaposhnikov.
\newblock {Higgs inflation at the critical point}.
\newblock {\em Phys. Lett.}, B734:249--254, 2014, 1403.6078.

\bibitem{Barbon:2009ya}
J.~L.~F. Barbon and J.~R. Espinosa.
\newblock {On the Naturalness of Higgs Inflation}.
\newblock {\em Phys. Rev.}, D79:081302, 2009, 0903.0355.

\bibitem{Burgess:2009ea}
C.~P. Burgess, Hyun~Min Lee, and Michael Trott.
\newblock {Power-counting and the Validity of the Classical Approximation
  During Inflation}.
\newblock {\em JHEP}, 09:103, 2009, 0902.4465.

\bibitem{Burgess:2010zq}
C.~P. Burgess, Hyun~Min Lee, and Michael Trott.
\newblock {Comment on Higgs Inflation and Naturalness}.
\newblock {\em JHEP}, 07:007, 2010, 1002.2730.

\bibitem{Ballesteros:2016xej}
Guillermo Ballesteros, Javier Redondo, Andreas Ringwald, and Carlos Tamarit.
\newblock {Standard Model—axion—seesaw—Higgs portal inflation. Five
  problems of particle physics and cosmology solved in one stroke}.
\newblock {\em JCAP}, 1708(08):001, 2017, 1610.01639.

\bibitem{Gerbino:2016sgw}
Martina Gerbino, Katherine Freese, Sunny Vagnozzi, Massimiliano Lattanzi, Olga
  Mena, Elena Giusarma, and Shirley Ho.
\newblock {Impact of neutrino properties on the estimation of inflationary
  parameters from current and future observations}.
\newblock {\em Phys. Rev.}, D95(4):043512, 2017, 1610.08830.

\bibitem{Abazajian:2016yjj}
Kevork~N. Abazajian et~al.
\newblock {CMB-S4 Science Book, First Edition}.
\newblock 2016, 1610.02743.

\bibitem{Matsumura:2013aja}
T.~Matsumura et~al.
\newblock {Mission design of LiteBIRD}.
\newblock 2013, 1311.2847.
\newblock [J. Low. Temp. Phys.176,733(2014)].

\bibitem{Alabidi:2012ex}
Laila Alabidi, Kazunori Kohri, Misao Sasaki, and Yuuiti Sendouda.
\newblock {Observable Spectra of Induced Gravitational Waves from Inflation}.
\newblock {\em JCAP}, 1209:017, 2012, 1203.4663.

\bibitem{Garcia-Bellido:2017aan}
Juan Garcia-Bellido, Marco Peloso, and Caner Unal.
\newblock {Gravitational Wave signatures of inflationary models from Primordial
  Black Hole Dark Matter}.
\newblock {\em JCAP}, 1709(09):013, 2017, 1707.02441.

\bibitem{Pattison:2017mbe}
Chris Pattison, Vincent Vennin, Hooshyar Assadullahi, and David Wands.
\newblock {Quantum diffusion during inflation and primordial black holes}.
\newblock {\em JCAP}, 1710(10):046, 2017, 1707.00537.

\end{thebibliography}
\bibliographystyle{hunsrt}

\end{document}